\documentclass[aps,noshowpacs,nofootinbib,superscriptaddress,notitlepage]{revtex4-1}
\usepackage{epsf} \usepackage{graphicx} \usepackage{amsmath}
\usepackage{hyperref}
\def\tstrut{\vrule height2.5ex depth0pt width0pt} 
\def\slashchar#1{\setbox0=\hbox{$#1$}
   \dimen0=\wd0 \setbox1=\hbox{/} \dimen1=\wd1
   \ifdim\dimen0>\dimen1 \rlap{\hbox to \dimen0{\hfil/\hfil}} #1
   \else  \rlap{\hbox to \dimen1{\hfil$#1$\hfil}} / \fi}

\begin{document}

\title{Photon emission in neutral current interactions at intermediate energies}
\author{E.~Wang}
\affiliation{Departamento de F\'\i sica Te\'orica and IFIC, Centro Mixto
Universidad de Valencia-CSIC, Institutos de Investigaci\'on de
Paterna, E-46071 Valencia, Spain} 
\author{L.~Alvarez-Ruso}
\affiliation{Instituto de F\'{\i}sica Corpuscular (IFIC), 
Centro Mixto CSIC-Universidad de Valencia,
Institutos de Investigaci\'on de Paterna, E-46071
Valencia, Spain}
\author{J.~Nieves}
\affiliation{Instituto de F\'{\i}sica Corpuscular (IFIC), 
Centro Mixto CSIC-Universidad de Valencia,
Institutos de Investigaci\'on de Paterna, E-46071
Valencia, Spain}


\begin{abstract}
Neutral current photon emission reactions 
with nucleons and nuclei are studied. These processes are important
backgrounds for $\nu_\mu \to \nu_e$ $(\bar\nu_\mu \to \bar \nu_e)$
appearance oscillation experiments where electromagnetic showers
instigated by electrons (positrons) and photons are not
distinguishable. At intermediate energies, these reactions are dominated
by the weak excitation of the $\Delta(1232)$ resonance and its
subsequent decay into $N\gamma$. There are also non-resonant
contributions that, close to threshold, are fully determined by the
effective chiral Lagrangian of strong interactions.  In addition, we have also
included mechanisms mediated by nucleon
excitations ($N^*$) from the second resonance region above the
$\Delta(1232)$. From these states, the contribution of the 
$D_{13}$ $N^*(1520)$ turns out to be sizable for (anti)neutrino
energies above 1.5 GeV. We have extended the model to nuclear targets
taking, into account Pauli blocking, Fermi motion and the in-medium
$\Delta$ resonance broadening. We present our predictions 
for both the incoherent and coherent channels, showing the relevance of the 
nuclear corrections. We also discuss the target mass dependence of the cross sections.   
This study is important in order
to reduce systematic effects in neutrino oscillation experiments.
\end{abstract}

\maketitle

\section{Introduction}

A good understanding of (anti)neutrino cross sections is
crucial to reduce the systematic uncertainties
in oscillation experiments aiming at a precise determination of
neutrino properties~\cite{Formaggio:2013kya}. Our present knowledge 
of neutrino-nucleus interactions  
has been significantly improved by a new generation of oscillation and cross section 
experiments. Quasielastic (QE) scattering measurements have been 
published by MiniBooNE~\cite{AguilarArevalo:2010zc,AguilarArevalo:2010cx,AguilarArevalo:2013hm} at 
neutrino energies $E_\nu \sim 1$~GeV, by MINER$\nu$A~\cite{Fields:2013zhk,Fiorentini:2013ezn} 
at $E_\nu \sim 3.5$~GeV and by NOMAD at high (3-100~GeV) energies~\cite{Lyubushkin:2008pe}. Detailed 
single pion production data have become available from 
MiniBooNE~\cite{AguilarArevalo:2009ww,AguilarArevalo:2010xt,AguilarArevalo:2010bm} for different reaction 
channels including the coherent one, which has also been studied by SciBooNE~\cite{Hiraide:2008eu,Kurimoto:2010rc} at $E_\nu \sim 1$~GeV 
and NOMAD~\cite{Kullenberg:2009pu}. Finally, new inclusive cross section results have been reported by T2K~\cite{Abe:2013jth}, SciBooNE~\cite{Nakajima:2010fp}, MINOS~\cite{Adamson:2009ju} and NOMAD~\cite{Wu:2007ab} Collaborations. These results challenge our understanding of neutrino interactions with matter and have triggered a renewed theoretical interest~\cite{Morfin:2012kn}. 
Quasielastic scattering has been investigated with a local Fermi gas~\cite{Graczyk:2003ru,Nieves:2004wx,Athar:2005hu,Martini:2009uj}, realistic spectral functions~\cite{Benhar:2005dj,Ankowski:2007uy}, different models to describe the interaction of the knocked-out nucleon with the residual nucleus~\cite{Martinez:2005xe,Butkevich:2007gm,Meucci:2011vd} and using the information from electron scattering data encoded in the scaling function~\cite{Caballero:2005sj}. The importance of two-nucleon contributions for the proper understanding of QE-like and inclusive cross sections has emerged in different studies~\cite{Martini:2009uj,Amaro:2010sd,Nieves:2011pp}, and  their impact in the kinematic neutrino-energy reconstruction has been stressed~\cite{Martini:2012fa,Nieves:2012yz,Lalakulich:2012hs}. Incoherent pion production has also been scrutinized using microscopic models for the reaction mechanism on the nucleon~\cite{Sato:2003rq,Hernandez:2007qq,Hernandez:2010bx,Leitner:2008ue,Serot:2012rd,Hernandez:2013jka}, with special attention paid to pion final state interactions in 
nuclei~\cite{Ahmad:2006cy,Leitner:2008wx,Golan:2012wx,Lalakulich:2012cj,Hernandez:2013jka}. New microscopic models have been 
developed for coherent pion production~\cite{Singh:2006bm,AlvarezRuso:2007tt,Amaro:2008hd,Nakamura:2009iq,Zhang:2012xi} while traditional ones, based on the partial conservation of the axial current (PCAC), have been updated~\cite{Paschos:2005km,Berger:2008xs,Hernandez:2009vm,Kopeliovich:2012tu}.

One of the possible reaction channels is photon emission induced by
neutral current (NC) interactions (NC$\gamma$), which can occur on single nucleons
and on nuclear targets.  Weak photon emission has a small cross
section compared, for example, with pion production, the most important
inelastic mechanism.  In spite of this, NC photon emission turns out
to be one of the largest backgrounds in $\nu_\mu \to \nu_e$
$(\bar\nu_\mu \to \bar \nu_e)$ oscillation experiments where electromagnetic
showers instigated by electrons (positrons) and photons are not
distinguishable. Thus, NC events producing single photons become an
irreducible background to the charge-current (CC) QE signatures of
$\nu_e$ ($\bar \nu_e$) appearance. This is precisely the case of the MiniBooNE experiment 
that was designed to test an earlier indication of a $\bar\nu_\mu \to \bar\nu_e$
oscillation signal observed at LSND~\cite{Athanassopoulos:1996jb,Athanassopoulos:1997pv}. 
The MiniBooNE experiment finds an excess of
events with respect to the predicted background in both $\nu$ and
$\bar \nu$ modes. In the $\bar \nu$ mode, the data are found to be
consistent with $\bar\nu_\mu \to \bar\nu_e$ oscillations and have some
overlap with the LSND result~\cite{Aguilar-Arevalo:2013pmq}. MiniBooNE 
data for $\nu_e$ appearance in the $\nu_\mu$ mode show a clear ($3 \sigma$) 
excess of signal-like events at low reconstructed neutrino energies  
($200 < E_\nu^{\mathrm{QE}} < 475$~MeV)~\cite{AguilarArevalo:2007it,Aguilar-Arevalo:2013pmq}. 
However, the $E_\nu^{\mathrm{QE}}$ distribution of the events is only marginally compatible 
with a simple two-neutrino oscillation model~\cite{Aguilar-Arevalo:2013pmq}. While
several exotic explanations for this excess have been proposed, it
could be related to unknown systematics or poorly understood backgrounds in the experimental analysis. 
In a similar way, NC$\gamma$ is a source of misidentified electron-like events in the $\nu_e$ 
appearance measurements at T2K~\cite{Abe:2013xua}. Even if the NC$\gamma$ contribution to the background 
is relatively small, it can be critical in measurements of the CP-violating phase.   
It is therefore very important to have a robust theoretical understanding of the NC photon emission reaction, 
which cannot be unambiguously constrained by data. This is the goal of the present work.

The first step forwards a realistic description of NC photon emission on nuclear targets of neutrino 
detectors is the study of the corresponding process on the nucleon. Theoretical models for the 
$\nu N \to \nu N \gamma$ reaction have been presented in Refs.~\cite{Hill:2009ek,Serot:2012rd}. They start 
from Lorentz-covariant effective field theories with nucleon, pion, $\Delta(1232)$ but also 
scalar ($\sigma$) and vector ($\rho$, $\omega$) mesons as the relevant degrees of freedom, and exhibit 
a nonlinear realization of (approximate) $SU(2)_L \otimes SU(2)_R$ chiral symmetry. The single mechanism of 
$\Delta(1232)$ excitation followed by its decay $\Delta \to N \gamma$ was considered in Ref.~\cite{Barbero:2012sb}, 
where a consistent treatment of the $\Delta$ vertices and propagator is adopted. 
Several features of the previous studies, 
in particular the approximate chiral symmetry and the dominance of the $\Delta(1232)$ mediated mechanism 
are common to the model derived in our work. In Ref.~\cite{Serot:2012rd}, a special attention is paid 
to the power counting, which is shown to be valid for neutrino energies below 550 MeV. However, 
the neutrino fluxes of most neutrino experiments span to considerably higher energies. 
Thus, in Ref.~\cite{Zhang:2012xn}, the power counting scheme was abandoned, and the 
model of \cite{Serot:2012rd} was phenomenologically extended to the
intermediate energies ($E_\nu \sim 1 $ GeV) relevant for the MiniBooNE $\nu$ flux, 
by including phenomenological form factors. Though the extension proposed for 
the $\Delta$ and the nucleon Compton-like mechanisms seems reasonable, the one 
for the contact terms notably increases the cross section above $\sim1$ GeV
(they are more significant for neutrinos than for antineutrinos). 
Since the contact terms and the associated form factors
are not well understood so far, the model predictions for $E_\nu > 1$~GeV
should be taken cautiously, as explicitly acknowledged in Ref.~\cite{Zhang:2012xn}.

In nuclear targets, the reaction can be incoherent when the final nucleus is 
excited (and fragmented) or coherent, when it remains in the ground state. 
It is also possible that, after nucleon knockout, the residual excited nucleus decays emitting 
low-energy $\gamma$ rays. This mechanism has been recently investigated~\cite{Ankowski:2011ei} 
and shall not be discussed here. The model of Ref~\cite{Hill:2009ek} has been applied to incoherent 
photon production in an impulse approximation that ignores nuclear corrections~\cite{Hill:2010zy}. 
These are also neglected in the coherent case, which is calculated by treating the nucleus as a 
scalar particle and introducing a form factor to ensure that the coherence is restricted to 
low-momentum transfers~\cite{Hill:2009ek}. More robust is the approach of 
Refs.~\cite{Zhang:2012aka,Zhang:2012xi} based on a chiral effective field theory for nuclei, again 
extended phenomenologically to higher energies~\cite{Zhang:2012xn}. 
In addition to  Pauli blocking and Fermi motion, the $\Delta$ resonance
broadening in the nucleus, is also taken into account. The latter correction causes a very strong 
reduction of the resonant contribution to the cross section, in variance with our results, as will 
be shown below. The ratio of the $\Delta$ to  photon and $\Delta$  to $\pi^0$ decay rates is enhanced 
in the nuclear medium by an amount that depends on the resonance invariant mass, momentum and 
also production 
position inside the nucleus, as estimated with a transport model~\cite{Leitner:2008fg,Leitner:2009zz}. 
The coherent channel has also been studied in Refs.~\cite{Gershtein:1980wu,Rein:1981ys} at high energies. 
A discussion about these works can be found in Section V.E of Ref.~\cite{Hill:2009ek}.
  
It is worth mentioning that both the models of Ref.~\cite{Hill:2009ek} and 
Refs~\cite{Serot:2012rd,Zhang:2012aka,Zhang:2012xi,Zhang:2012xn} have been used to calculate the 
NC$\gamma$ events at MiniBooNE with contradicting conclusions~\cite{Hill:2010zy,Zhang:2012xn}. 
While in Ref.~\cite{Hill:2010zy} the number of these events were calculated to be twice as many as expected 
from the MiniBooNE in situ estimate, much closer values were predicted in Ref.~\cite{Zhang:2012xn}. 
The result that NC$\gamma$ events give a significant contribution to the MiniBooNE low-energy excess~\cite{Hill:2009ek} 
could have its origin in the lack of nuclear effects and rather strong detection efficiency correction.

Here we present a realistic model for NC photon emission in the $E_\nu \sim 1$~GeV 
region that extends and improves certain relevant aspects of the existing descriptions. 
The model is developed in the line of previous work on weak pion production on 
nucleons~\cite{Hernandez:2007qq} and nuclei for both incoherent~\cite{Hernandez:2013jka} 
and coherent~\cite{AlvarezRuso:2007it,Amaro:2008hd} processes.  
The model for free nucleons satisfies the approximate chiral symmetry
incorporated in the non-resonant terms and includes the dominant 
$\Delta(1232)$ excitation mechanism, with couplings and form factors taken from the available 
phenomenology. Moreover, we have extended the validity of the
approach to higher energies by including intermediate excited states from the second 
resonance region [$P_{11}(1440)$, $D_{13}(1520)$ and $S_{11}(1535)$]. Among them, we have 
found a considerable contribution of the $D_{13}(1520)$ for (anti)neutrino energies above 1.5 GeV. 
When the reaction takes place inside the nucleus, we have applied a series of standard 
medium corrections that have been extensively confronted with experiment in similar
processes such as pion~\cite{Oset:1981ih,Nieves:1991ye}, 
photon~\cite{Carrasco:1989vq} and electron~\cite{Gil:1997bm,Gil:1997jg} scattering with nuclei, or 
coherent pion photo~\cite{Carrasco:1991we} and electroproduction~\cite{Hirenzaki:1993jc}.

 This paper is organized as follows.  The model for NC production of
 photons off nucleons is described in Sec.~\ref{sec:nucleon}. After
 discussing the relevant kinematics, we evaluate the different
 amplitudes in Subsec.~\ref{sec:gamma_amp}. In the first place, the
 dominant $\Delta(1232)$ and non-resonant contributions are studied
 (Subsec.~\ref{sec:delta-bkg}). Next, we examine the contributions
 driven by $N^*$ resonances from the second resonance region
 (Subsec.~\ref{sec:secondregion}).  The relations between vector form
 factors and helicity amplitudes, and the off-diagonal $N^*N$
 Goldberger-Treiman (GT) relations are discussed in Appendices
 \ref{sec:heliamp} and \ref{sec:axialcoupling}, respectively.
 NC$\gamma$ reactions in nuclei are studied in
 Sec.~\ref{sec:ncgnuclei}. First, in
 Subsec.~\ref{sec:1p1h}, we pay attention to the incoherent channel driven by one
 particle--one hole (1p1h) nuclear excitations. Next, in
 Subsec.~\ref{sec:coherent}, the coherent channel is studied. We present our results in 
 Sec.~\ref{sec:results}, where we also compare some of our predictions
 with the corresponding ones from Refs.~\cite{Hill:2009ek,Zhang:2012xn}. This Section
 is split in two Subsections,  where  the results for NC$\gamma$ on
 single nucleons (Subsec.~\ref{sec:res1}) and on nuclei
 (Subsec.~\ref{sec:res2}) are discussed. Predictions for nuclear
 incoherent and coherent reactions are presented in
 Subsecs.~\ref{sec:res2a} and \ref{sec:res2b}, respectively. Finally
 the main conclusions of this work are summarized in Sec.~\ref{sec:concl}.

\section{Neutral current photon emission off nucleons}
\label{sec:nucleon}
In this section, we describe the model for  NC production of photons off nucleons,
\begin{equation}
\nu (k)  N (p) \to  \nu (k') N (p') \gamma (k_\gamma), \qquad
\bar\nu (k)  N (p) \to  \bar\nu (k') N (p') \gamma (k_\gamma)  \label{eq:neu}
\end{equation}

\subsection{Kinematics and general definitions}
\begin{figure}[tbh]
\centerline{\includegraphics[width=0.55\textwidth]{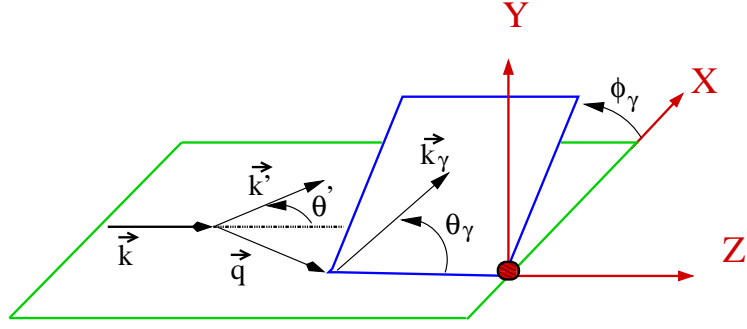}}
\caption{Representation of the different LAB kinematical variables
used through this work.}\label{fig:coor}
\end{figure}
The unpolarized differential cross section with respect to the
photon kinematical variables (kinematics is sketched in Fig.~\ref{fig:coor}) is given in the
Laboratory (LAB) frame by
\begin{equation}
\frac{d^{\,3}\sigma_{(\nu,\bar\nu)}}{dE_\gamma d\Omega(\hat{k}_\gamma)} =
    \frac{E_\gamma}{ |\vec{k}|}\frac{G^2}{16\pi^2}
     \int  \frac{d^3k'}{|\vec{k}^{\prime}\,|}
    L_{\mu\sigma}^{(\nu,\bar\nu)}W^{\mu\sigma}_{{\rm NC}\gamma} \label{eq:sec} \,.
\end{equation}
As we neglect the neutrino masses, $E_\nu = |\vec{k}|$, $E^{\prime} = |\vec{k}^{\prime}\,|$ and $E_\gamma=  |\vec{k}_\gamma|$, where $\vec{k}$, $\vec{k}^{\prime}$ and $\vec{k}_\gamma$ are the incoming neutrino, outgoing neutrino and outgoing photon momenta in LAB, in this order; $G=1.1664\times 10^{-11}$ MeV$^{-2}$ is the Fermi constant, while $L ^{(\nu,\bar\nu)}$ and $W_{{\rm NC}\gamma}$ stand for the leptonic and hadronic tensors, respectively. The leptonic tensor\footnote{Our conventions are such that $\epsilon_{0123}= +1$ and $g^{\mu\nu}=(+,-,-,-)$.}  
\begin{eqnarray}
L_{\mu\sigma}^{(\nu,\bar\nu)}&=& (L_s)_{\mu\sigma}+ i
 (L^{(\nu,\bar \nu)}_a)_{\mu\sigma} =
 k^\prime_\mu k_\sigma +k^\prime_\sigma k_\mu
+ g_{\mu\sigma} \frac{q^2}2 \pm i
\epsilon_{\mu\sigma\alpha\beta}k^{\prime\alpha}k^\beta, \qquad \left(+\to
\nu, \, -\to \bar\nu \right) \label{eq:lep} \,,
\end{eqnarray}
is orthogonal to the four momentum transfer $q_\mu=k_\mu-k'_\mu$, with $q^2=-2k\cdot k'=-4EE'\sin^2\theta'/2$. The hadronic tensor
includes the non-leptonic vertices and reads
\begin{eqnarray}
W^{\mu\sigma}_{{\rm NC} \gamma} &=& \frac{1}{4M}\overline{\sum_{\rm
 spins}} \int\frac{d^3p^\prime}{(2\pi)^3} \frac{1}{2E^\prime_N}
  \delta^4(p^\prime+k_\gamma-q-p) \langle N \gamma |
 j^\mu_{\rm NC\gamma}(0) | N \rangle \langle N \gamma | j^\sigma_{\rm NC \gamma}(0) | N
 \rangle^* \label{eq:wmunu-nucleon} \,,
\end{eqnarray}
with $M$ the nucleon mass\footnote{We take the average of the neutron
and proton masses.} and $E^\prime_N$ the energy of the
outgoing nucleon.  The bar over the sum of initial and final spins
denotes the average over the initial ones. The one particle
states are normalized as $\langle \vec{p}\, |
\vec{p}^{\,\prime} \rangle = (2\pi)^3 2p_0
\delta^3(\vec{p}-\vec{p}^{\,\prime})$. Then, the matrix element  
$\langle N \gamma | j^\mu_{\rm NC\gamma}(0) | N \rangle$ is dimensionless. 
For the sake of completeness, we notice that the NC, $j^\mu_{\rm NC}$  and electromagnetic (EM), $s^\mu_{\rm EM}$
currents at the quark level are given by  
\begin{eqnarray}
j^\mu_{\rm NC} &=& \bar{\Psi}_u\gamma^\mu(1-\frac83 \sin^2\theta_W-
\gamma_5)\Psi_u  - \bar{\Psi}_d\gamma^\mu(1-\frac43 \sin^2\theta_W-
\gamma_5)\Psi_d  - \bar{\Psi}_s\gamma^\mu(1-\frac43 \sin^2\theta_W-
\gamma_5)\Psi_s  \,, \nonumber\\
&=&\bar\Psi_q \gamma^\mu (1-\gamma_5) \tau^{(1)}_0 \Psi_q
- 4 \sin^2 \theta_W s^\mu_{\rm EM}
-\bar\Psi_s\gamma^\mu(1-\gamma_5)\Psi_s \,, \\
s^\mu_{\rm EM} &=& \frac23 \bar{\Psi}_u\gamma^\mu \Psi_u 
- \frac13 \bar{\Psi}_d\gamma^\mu \Psi_d 
- \frac13 \bar{\Psi}_s\gamma^\mu \Psi_s \,,
\end{eqnarray}
where $\Psi_u$, $\Psi_d$ and $\Psi_s$ are the quark fields and $\theta_W$ 
the weak angle ($\sin^2\theta_W= 0.231$). The zeroth spherical component of 
the isovector operator $\tau^{(1)}$ is equal to the third component of the 
isospin Pauli matrices $\vec{\tau}$. 

By construction, the hadronic tensor accomplishes
\begin{eqnarray}
W^{\mu\sigma}_{{\rm NC} \gamma}= 
W^{(s)\mu\sigma}_{{\rm NC} \gamma} + i
W^{(a)\mu\sigma}_{{\rm NC} \gamma} \,,
\end{eqnarray}
in terms of its real symmetric, $W^{(s)}_{{\rm NC} \gamma}$, and antisymmetric, 
$W^{(a)}_{{\rm NC} \gamma}$, parts.  Both lepton and hadron tensors are independent
of the neutrino flavor and, therefore, the cross section for the
reaction of Eq.~(\ref{eq:neu}) is the same for electron, muon or tau
incident (anti)neutrinos.

Let us define the amputated amplitudes $\Gamma^{\mu \rho}$, as
\begin{equation}
\langle N \gamma |
 j^\mu_{\rm NC\gamma}(0) | N \rangle = \bar u(p') \Gamma^{\mu \rho} u (p) \epsilon^*_\rho(k_\gamma) \,,
\end{equation}
where the spin dependence of the Dirac spinors (normalized such that
$\bar uu= 2M$) for the nucleons is understood, and
$\epsilon(k_\gamma)$ is the polarization vector of the outgoing
photon. To keep the notation simple we do not specify the type of nucleon 
($N=n$ or $p$) in $\Gamma^{\mu \rho}$. In terms of these
amputated amplitudes, and after performing the average (sum) over the
initial (final) spin states, we find
\begin{eqnarray}
W^{\mu\sigma}_{{\rm NC} \gamma} &=&- \frac{1}{8M} \int\frac{d^3p^\prime}{(2\pi)^3} \frac{1}{2E^\prime_N}
  \delta^4(p^\prime+k_\gamma-q-p) {\rm Tr}\left[
  (\slashchar{p}'+M)\Gamma^{\mu\rho}(\slashchar{p}+M) \gamma^0
  (\Gamma^\sigma_{.\,\rho})^\dagger \gamma^0 \right] \,.
\label{eq:wmunu-nucleon-avg}
\end{eqnarray}
After performing the $d^3p^\prime$ integration, 
there is still a $\delta(p^{\prime\,0}+E_\gamma-q^0-p^0) $ 
left in the hadronic tensor, which can be used to perform the integration over $| \vec{k}^\prime |$ 
in Eq.~(\ref{eq:sec}).

\subsection{Evaluation of the $\Gamma^{\mu \rho}$ amputated amplitudes}
\label{sec:gamma_amp}
\subsubsection{The $\Delta(1232)$ contribution, chiral symmetry and non-resonant terms}
\label{sec:delta-bkg}

\begin{figure}[ht!]
    \begin{center}
           \includegraphics[width=0.52\textwidth]{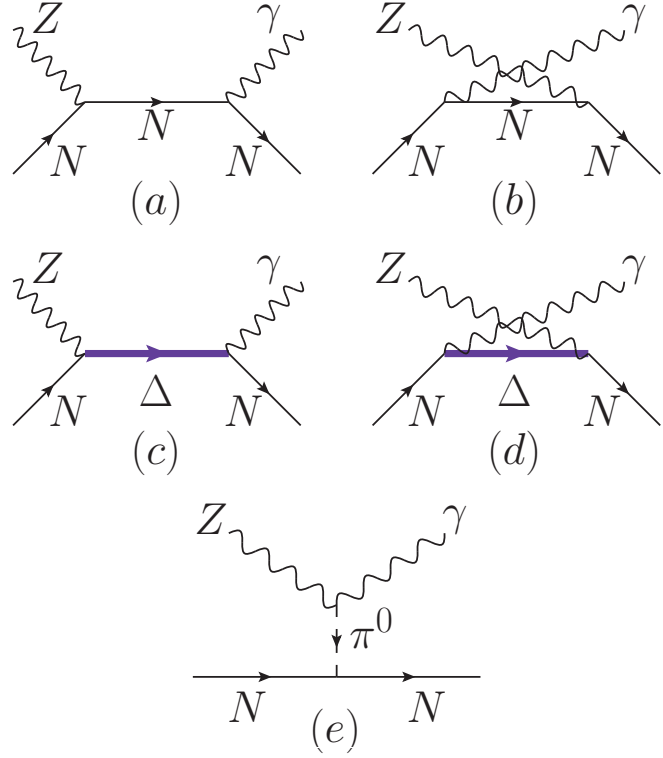}
            \caption{Model for photon emission off the
              nucleon; direct and crossed nucleon-pole terms (a,b), direct and crossed
              $\Delta(1232)$-pole terms (c,d) and the anomalous
              $t-$channel pion exchange term (e). Throughout this work, we
              denote these contributions as $NP$, $CNP$, $\Delta
              P$, $C\Delta P$ and $\pi Ex$, respectively.
    \label{fig:five}}
    \end{center}
\end{figure}

Just as in pion production~\cite{Hernandez:2007qq}, one expects the NC$\gamma$ reaction to be dominated by 
the excitation of the $\Delta(1232)$ supplemented with a non-resonant background. In our case, the leading 
non-resonant contributions are nucleon-pole terms built out of  $Z^0NN$ and $\gamma NN$ vertices that 
respect chiral symmetry. The $q^2$ dependence of the amplitudes is introduced via phenomenological form factors.
We also take into account the subleading mechanism originated from the anomalous $Z^0\gamma\pi$ vertex, that involves
a pion exchange in the $t-$channel. Thus, in a first stage we consider the five diagrams depicted in Fig.~\ref{fig:five}.  
 The corresponding amputated amplitudes are
\begin{eqnarray}
\Gamma^{\mu \rho}_{N} = \Gamma^{\mu \rho}_{NP} + \Gamma^{\mu
  \rho}_{CNP} &=& i e\,
J^\rho_{EM}(-k_\gamma)\frac{\slashchar{p}+\slashchar{q}+M}{(p+q)^2-M^2
+i\epsilon} J^\mu_{NC}(q)   
 +
i e\,  J^\mu_{NC}(q)\frac{(\slashchar{p}'-\slashchar{q}+M)}{(p'-q)^2-M^2+i\epsilon} 
J^\rho_{EM}(-k_\gamma) \label{eq:NP-CNP} \,, \\
\Gamma^{\mu \rho}_{\pi Ex} & = &  eC_N\frac{g_A M}{4\pi^2 f_\pi^2}(1-4
\sin^2\theta_W) \frac{\epsilon^{\mu\rho\sigma\alpha}q_\sigma (k_{\gamma})_\alpha}{(q-k_\gamma)^2-m_\pi^2} \gamma_5, \qquad \left(C_N=+1\to
p, \, C_N=-1 \to n \right) \\
\Gamma^{\mu \rho}_{\Delta} = \Gamma^{\mu \rho}_{\Delta P} + \Gamma^{\mu
  \rho}_{C\Delta P} &=& i e\,
\gamma^0 \left[ J^{\alpha \rho}_{EM}(p',k_\gamma)\right]^\dagger
\gamma^0\frac{P_{\alpha\beta}
  (p+q)}{(p+q)^2-M_\Delta^2
+i M_\Delta \Gamma_\Delta} J^{\beta \mu}_{NC}(p,q) \nonumber \\   
&+&
i e\, \gamma^0 \left[  J^{\alpha \mu}_{NC}(p',-q) \right]^\dagger
\gamma^0 \frac{P_{\alpha\beta}
  (p^{\,\prime}-q)}{(p^{\,\prime}-q)^2-M_\Delta^2
+i \epsilon}J^{\beta \rho}_{EM}(p,-k_\gamma) \label{eq:DP-CDP} \,,
\end{eqnarray}
with $e > 0$ the electron charge,  such that $\alpha=e^2/4\pi \approx
1/137$, $f_\pi = 92.4$ MeV the pion decay constant and $g_A=1.267$
the axial nucleon charge; $m_\pi$ and 
$M_\Delta (\sim 1232\, {\rm MeV})$ are the pion and $\Delta$ masses, respectively.
As it will be clear in the following, each of the building blocks of the model
is gauge invariant by construction $ \bar u (p') \, \Gamma_{N,\Delta,\pi Ex}^{\mu \rho} \, u(p) (k_\gamma)_\rho = 0$. 
The vector parts of these amplitudes are also conserved (CVC)
$ \bar u (p') \, V_{N,\Delta,\pi Ex}^{\mu \rho} \, u(p) q_\mu = 0$.

\paragraph{$NP$ and $CNP$ amplitudes:}

The nucleon NC and EM currents are given by
\begin{eqnarray}
J^\mu_{NC}(q)&=&\gamma^\mu \tilde{F}_1(q^2)+\frac{i}{2M}\sigma^{\mu\beta}q_\beta \tilde{F}_2(q^2) - \gamma^\mu \gamma_5 \tilde{F}_A(q^2),   \\
J^\mu_{EM}(k_\gamma)&=& \gamma^\mu F_1(0) + \frac{i}{2M}\sigma^{\mu\nu}
(k_{\gamma})_\nu F_2(0),
\end{eqnarray}
where $\tilde{F}_{1,2}$ and $\tilde{F}_A$ are the NC vector and axial form factors\footnote{Note that pseudoscalar ($q^\mu \gamma_5$) terms do not contribute because $q^\mu L_{\mu\sigma}^{(\nu\bar\nu)}=0$ when neutrino masses are neglected.} while  $F_{1,2}$ are the EM ones. 
These form factors take different values for protons and neutrons. For $F_{1,2}$, we have that  
\begin{eqnarray}
F^{(N)}_1 &=& \frac{G^N_E+\tau G^N_M}{1+\tau},\quad F^{(N)}_2 = \frac{G^N_M -
  G^N_E}{1+\tau}, \qquad N=p,n
\end{eqnarray}
with
\begin{equation}
 G_E^p = \frac{G_M^p}{\mu_p}=
\frac{G_M^n}{\mu_n} = -(1+ b \tau) \frac{G_E^n}{\mu_n  a \tau} =
\left(\frac{1}{1-q^2/M^2_D}\right)^2 \,,
\end{equation}
where $\tau = -q^2/4M^2$, $M_D=0.84$ GeV, $\mu_p=2.793$, $\mu_n=-1.913$, $b=4.61$ and $a=0.942$~\cite{Krutov:2002tp}.
 
The NC vector form factors $\tilde{F}_{1,2}$ can be referred to the
EM ones thanks to isospin symmetry relationships,
\begin{eqnarray}
 \tilde{F}^{(p)}_{1,2}&=&(1-4 \sin^2\theta_W) F^{(p)}_{1,2} -F^{(n)}_{1,2} -F^{(s)}_{1,2}\\
 \tilde{F}^{(n)}_{1,2}&=&(1-4 \sin^2\theta_W) F^{(n)}_{1,2} -F^{(p)}_{1,2} -F^{(s)}_{1,2} \,,
\end{eqnarray}
where $F^{(s)}_{1,2}$ are the strange EM form factors. Furthermore, in the axial sector one has that
\begin{equation}
 \tilde{F}^{(p,n)}_A =  \pm F_A  - F^{(s)}_A, \qquad \left(+\to
p, \, -\to n\right) \,,
\end{equation}
where $F_A$ is the axial form factor that appears in CCQE interactions, for which we adopt 
a conventional dipole parametrization
\begin{equation}
F_A(q^2) = g_A \left(1-\frac{q^2}{M^2_A} \right)^{-2} \label{eq:gaxial}
\end{equation}
with an  axial mass $M_A=1$~GeV~\cite{Bodek:2007ym}; $F^{(s)}_A$ is the strange axial form factor. 
At present, the best determinations of the strange form factors are consistent with zero~\cite{Pate:2013wra}, thus 
they have been neglected in the present study. 
 
\paragraph{$\pi Ex$ amplitudes:}

The $t-$channel pion exchange contribution arises from the anomalous
($\pi^0\gamma Z^0$) Lagrangian~\cite{Hill:2009ek}
\begin{equation}
{\cal L}_{\pi^0\gamma Z^0}= \frac{eg}{4\cos\theta_W}
\frac{N_C}{12\pi^2f_\pi}(1-4 \sin^2\theta_W) \pi^0
\epsilon^{\mu\nu\alpha\beta}\partial_\mu Z_\nu\partial_\alpha
A_\beta \label{eq:WZW}
\end{equation}
together with the leading order $\pi^0NN$ interaction term
\begin{equation} 
{\cal L}_ {\rm
\pi^0NN}= \frac{g_A}{f_\pi} \bar\Psi \gamma^\mu\gamma_5
\frac{\tau_3}{2}(\partial_\mu \pi^0)\Psi \,, \qquad 
\Psi= \left (\begin{array}{c}p\cr n\end{array}\right ) \,,
\end{equation}
where $\Psi$, $\pi^0$, $A_\beta$, $Z_\nu$ are the nucleon, neutral pion, 
photon and $Z^0$ boson fields, respectively.
Besides, $g = e/\sin \theta_W$ is related to the Fermi
constant $G$ and the $W$-boson mass as $G/\sqrt 2 = g^2/8M^2_W$; 
$N_C$ is the number of colors. 
The Lagrangian of Eq.~(\ref{eq:WZW})
arises from the Wess-Zumino-Witten
term~\cite{Wess:1971yu,Witten:1983tw}, which accounts for the axial anomaly 
of QCD.  

\paragraph{$\Delta P$ and $C \Delta P$ amplitudes:}

In the $\Delta-$driven amplitudes of Eq.~(\ref{eq:DP-CDP}), $P^{\mu\nu}$ is the spin
3/2 projection operator, which reads
\begin{equation}
P^{\mu\nu}(p_\Delta)= - (\slashchar{p}_\Delta + M_\Delta) \left [ g^{\mu\nu}-
  \frac13 \gamma^\mu\gamma^\nu-\frac23\frac{p_\Delta^\mu
  p_\Delta^\nu}{M_\Delta^2}+ \frac13\frac{p_\Delta^\mu
  \gamma^\nu-p_\Delta^\nu \gamma^\mu }{M_\Delta}\right] \,;  \label{eq:spin32}
\end{equation}
$\Gamma_\Delta$ is the resonance width in its rest frame, given by
\begin{equation}
\Gamma_\Delta(s) = \frac{1}{6\pi} \left ( \frac{f^*}{m_\pi}\right )^2
 \frac{M}{\sqrt s} \left [ \frac{\lambda^\frac12
  (s,m_\pi^2,M^2)}{2\sqrt s} \right ]^3 \Theta(\sqrt s
-M-m_\pi),\qquad s= p_\Delta^2 ,
\end{equation}
with $f^*=2.14$, the $\pi N \Delta$ coupling obtained from the empirical $\Delta \to N \pi$ decay width (see Table~\ref{tab:resonances}); 
$\lambda(x,y,z) = x^2+y^2+z^2-2xy-2xz-2yz$, and $\Theta$ is the step function.

 The weak NC and EM currents for the nucleon to $\Delta$  transition are the
 same for protons or neutrons and  are given by
\begin{eqnarray}
\frac12 J^{\beta\mu}_{NC}(p,q) &=& \left[ \frac{\tilde{C}^V_3(q^2)}{M} (g^{\beta\mu} \slashchar{q}
    -q^\beta \gamma^\mu ) +\frac{\tilde{C}^V_4(q^2)}{M^2} (g^{\beta\mu} q \cdot p_\Delta
    -q^\beta p^\mu_\Delta )+\frac{\tilde{C}^V_5(q^2)}{M^2} (g^{\beta\mu} q \cdot p
    -q^\beta p^\mu ) \right] \gamma_5 \nonumber\\
 &+&\frac{\tilde{C}^A_3(q^2)}{M} (g^{\beta\mu}\slashchar{q}-q^\beta \gamma^\mu ) 
   + \frac{\tilde{C}^A_4(q^2)}{M^2} (g^{\beta\mu} q \cdot p_\Delta
    -q^\beta p^\mu_\Delta ) + \tilde{C}^A_5(q^2) g^{\beta\mu}\label{eq:nc}\,,
\end{eqnarray}
\begin{eqnarray}
J^{\beta\rho}_{EM}(p,-k_\gamma) &=& -\left[ \frac{C^V_3(0)}{M}
  (g^{\beta\rho} \slashchar{k}_\gamma-k_\gamma^{\beta} \gamma^\rho )
  +\frac{C^V_4(0)}{M^2} (g^{\beta\rho} k_\gamma \cdot p_{\Delta c}
  -k_\gamma^\beta p^\rho_{\Delta c} )
+ \frac{C^V_5(0)}{M^2} (g^{\beta\rho} k_\gamma \cdot p -k_\gamma^\beta p^\rho ) \right] \gamma_5 \,,
\end{eqnarray}
where $p_\Delta=p+q$ and $p_{\Delta c}= p-k_\gamma$; $\tilde{C}^V_i$, $\tilde{C}^A_i$ and $C^V_i$ are the NC vector, NC axial\footnote{There is another contribution to the axial current
  $\tilde{C}^A_6(q^2) q^\beta q^\mu$, which does not contribute to the cross
  section because $q^\mu L_{\mu\sigma}^{(\nu\bar\nu)}=0$ for
  massless neutrinos.}
  and EM transition form factors, respectively. As in the nucleon case, the NC vector form factors are related  to the EM ones 
\begin{equation}
\tilde{C}^V_i(q^2) = (1-2 \sin^2 \theta_W) C^V_i(q^2)
\end{equation}
according to the isovector character of the $N-\Delta$ transition. These EM form factors (and couplings) can be constrained 
using experimental results on pion photo and electroproduction in the $\Delta$ resonance region. In particular, 
they can be related to the helicity amplitudes  $A_{1/2}$, $A_{3/2}$ and $S_{1/2}$~\cite{Lalakulich:2006sw,Leitner:2008ue} 
commonly extracted in the analyses of meson electroproduction data. The explicit expressions are given in Appendix~\ref{sec:heliamp}. For the helicity amplitudes and their $q^2$ dependence we have taken the parametrizations of the MAID analysis~\cite{Drechsel:2007if,MAID}.~\footnote{The set of $N-\Delta(1232)$ vector form factors used in \cite{Hernandez:2007qq}, which were taken from Ref.~\cite{Lalakulich:2006sw}, lead to negligible changes in the results compared to those presented below.} In the axial sector, we adopt the Adler model~\cite{Adler:1968tw,Bijtebier:1970ku}
\begin{equation}
\tilde{C}^A_3(q^2) = 0,\qquad \tilde{C}^A_4(q^2) = -\frac{\tilde{C}^A_5(q^2)}{4}\,,
\end{equation}
for the subleading (in a $q^2$ expansion) form factors and assume a standard dipole for the dominant 
\begin{equation} 
\tilde{C}^A_5(q^2) = C^A_5(0) \left(1-\frac{q^2}{M^2_A} \right)^{-2} \,,
\label{eq:c5} 
\end{equation}
 with  $C^A_5(0)=1.00\pm0.11$ and $M_A=0.93$~GeV fixed in a fit to $\nu_\mu d \to \mu^- \Delta^{++} n$ BNL and ANL data~\cite{Hernandez:2010bx}.

\subsubsection{The second resonance region}
\label{sec:secondregion}
Here, we extend the formalism to the second resonance region, 
which includes three isospin 1/2 baryon resonances
$P_{11}(1440)$, $D_{13}(1520)$ and $S_{11}(1535)$ (see Table \ref{tab:resonances}). 
In this way, we extend the validity of the model to higher energies. 
A basic problem that has to be faced with resonances is the determination 
of the transition form factors (coupling constants and $q^2$ dependence). 
As for the $\Delta(1232)$, we obtain vector form factors 
from the helicity amplitudes parametrized in Ref.~\cite{Drechsel:2007if}. 
The equations relating helicity amplitudes and form factors are 
compiled in Appendix~\ref{sec:heliamp}. Our knowledge of the axial transition 
form factors is much poorer. Some constraints can
be imposed from PCAC and the 
pion-pole dominance of the pseudoscalar form factors. These allow to 
derive off-diagonal Goldberger-Treiman (GT) relations between the 
leading axial couplings and the $N^* \to N \pi$ partial decay widths 
(see Table~\ref{tab:resonances} and Appendix~\ref{sec:axialcoupling} for more details). 
\begin{table}[htpb]
\caption{Properties of the resonances included in our model~\cite{Beringer:1900zz}. For each
  state, we list the Breit-Wigner mass ($M_R$) , spin ($J$),
  isospin ($I$), parity ($P$), total decay width
  ($\Gamma$), and axial coupling (denoted $F_A(0)$ for
  spin 1/2 states and $C^A_5(0)$ for spin 3/2 states).} 
\label{tab:resonances} \vspace{0.1cm}
\begin{center}
\begin{tabular}{c|ccccccc}
\hline\hline
 & $M_R$ [MeV] &$J$ &$I$ &~$P$~ &$\Gamma$ [MeV] & $\Gamma(R \to N \pi)/\Gamma$ & $F_A(0)$ or $C^A_5(0)$ \\ \hline
$\Delta(1232)$ &1232 &3/2 &3/2 &+ &117 & 100\% &$1.00\pm
0.11$~\footnote{In the case of the 
  $\Delta$, we use a $C^A_5(0)$ value obtained 
 in a reanalysis~\cite{Hernandez:2010bx} of the $\nu_\mu p \to \mu^- p \pi^+$ ANL and
  BNL bubble chamber data, which is smaller than 
the corresponding GT relation by $\sim 20$\%.}  
\\
$N(1440)$ &1440 &1/2 &1/2 &+ &300 & 65\% & $-$0.47 \\  
$N(1520)$ &1520 &3/2 &1/2 &$-$ &  115 & 60\%&$-$2.14 \\
$N(1535)$ &1535 &1/2 &1/2 &$-$ &  150 & 45\%&$-$0.21 \\
\hline\hline
\end{tabular}\end{center}
\end{table}

\begin{figure}[h!]
\begin{center}
\includegraphics[width=0.52\textwidth]{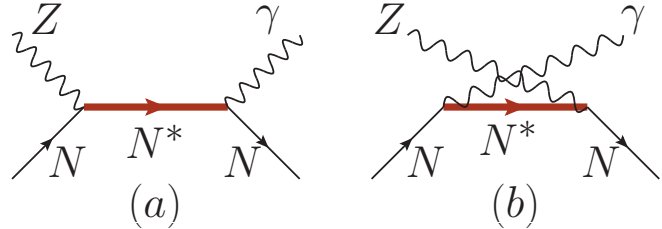}      
\caption{Direct (a) and crossed (b) $N^*$ pole contributions to the NC photon
emission process. We have considered the three resonances $\left[N(1440),
  N(1535), N(1520)\right]$ right above the $\Delta(1232).$} \label{fig:highR}
\end{center}
\end{figure}

For each of the three $P_{11}(1440)$, $D_{13}(1520)$ and
$S_{11}(1535)$ states, we have considered the contribution of direct ($RP$) and
crossed ($CRP$) resonance pole terms as depicted in Fig.~\ref{fig:highR}.  

\paragraph{$N(1440)$ and $N(1535)$:} 

The structure of the contribution of these two resonances to the amputated amplitudes is
similar to the one of the nucleon [Eq.~(\ref{eq:NP-CNP})]. We have
\begin{eqnarray}
\Gamma^{\mu \rho}_{R} = \Gamma^{\mu \rho}_{RP} + \Gamma^{\mu
  \rho}_{CRP} &=& ie\,
J^\rho_{EM(R)}(-k_\gamma)\frac{\slashchar{p}+\slashchar{q}+M_R}{(p+q)^2-M_R^2
+iM_R \Gamma_R} J^\mu_{NC(R)}(q) \nonumber  \\
& + &
ie\,  J^\mu_{NC(R)}(q)\frac{(\slashchar{p}'-\slashchar{q}+M_R)}{(p'-q)^2-M^2_R+i \epsilon} 
J^\rho_{EM(R)}(-k_\gamma) \,;
\end{eqnarray}
the resonance masses $M_R$ are listed in Table~\ref{tab:resonances} while the widths $\Gamma_R$ 
are discussed in Appendix~\ref{sec:width}. The EM and NC currents read
\begin{eqnarray}
  \nonumber \\
J^\mu_{NC(P_{11})}(q) &=&
\frac{\tilde{F}_{1(P_{11})}(q^2)}{(2M)^2}(\slashchar{q} q^\mu-q^2\gamma^\mu) +
\frac{\tilde{F}_{2(P_{11})}(q^2)}{2M}i\sigma^{\mu\nu}q_{\nu} +\tilde{F}_{A(P_{11})} (q^2) \gamma^\mu \gamma_5 \,,
\\
J^\mu_{EM(P_{11})}(k_\gamma)&=&\frac{F_{1(P_{11})}(0)}{(2M)^2} \slashchar{k}_\gamma k_\gamma^\mu + \frac{F_{2(P_{11})}(0)}{2M}i\sigma^{\mu\nu}(k_{\gamma})_\nu 
\end{eqnarray}
for the $N(1440)$ and
\begin{eqnarray}
J^\mu_{NC(S_{11})}(q) &=&\left[
  \frac{\tilde{F}_{1(S_{11})}(q^2)}{(2M)^2}(\slashchar{q}
  q^\mu-q^2\gamma^\mu) +
  \frac{\tilde{F}_{2(S_{11})}(q^2)}{2M}i\sigma^{\mu\nu}q_{\nu} \right] \gamma_5 
  +\tilde{F}_{A(S_{11})} (q^2)\gamma^\mu \,,
\\
J^\mu_{EM (S_{11})}(k_\gamma)&=&\left[ \frac{F_{1(S_{11})}(0)}{(2M)^2} \slashchar{k}_\gamma k_\gamma^\mu  + \frac{F_{2(S_{11})}(0)}{2M}i\sigma^{\mu\nu}(k_{\gamma})_\nu \right] \gamma_5 
\end{eqnarray}
for the $N(1535)$.~\footnote{Note that by construction gauge
  invariance and CVC are satisfied. This is also the case for the
  $N(1520)$ amplitudes that will be discussed next.} 
As in the nucleon case, isospin symmetry implies that 
\begin{eqnarray}
 \tilde{F}^{(p)}_{1,2(R)}&=&\left(1-4 \sin^2\theta_W \right)
F^{(p)}_{1,2(R)} -F^{(n)}_{1,2(R)} -F^{(s)}_{1,2(R)} \,, \nonumber \\
 \tilde{F}^{(n)}_{1,2(R)}&=& \left(1 -4 \sin^2\theta_W \right)
F^{(n)}_{1,2(R)} -F^{(p)}_{1,2(R)}-F^{(s)}_{1,2(R)}\,, 
\end{eqnarray}
with $F^{(N)}_{1,2 (P_{11},S_{11})}$ expressed in terms of the corresponding 
helicity amplitudes (see Appendix~\ref{sec:heliamp}). The NC axial 
form factors are 
\begin{eqnarray} 
 \tilde{F}^{(p,n)}_{A(R)} &=& \pm  F_{A(R)} + F^s_{A(R)}, \qquad \left(+\to
p, \, -\to n\right) \nonumber \\
F_{A(R)}(q^2) &=& F_{A(R)}(0) \left(1-\frac{q^2}{M^{* 2}_A} \right)^{-2}\label{eq:faj12t12} \,.
\end{eqnarray}
The couplings $F_{A(P_{11},S_{11})}(0)$ are obtained from the GT corresponding relations and have values 
given in Table~\ref{tab:resonances}. The $q^2$ dependence of these form factors is unknown so we have 
assumed a dipole ansatz with a natural value of  $M^*_A=1.0$ GeV for the axial mass. 
No information is available about the strange form factors $F^{(s)}_{1,2,A (P_{11},S_{11})}$
but they are likely to be small and to have a negligible impact on the 
observables, so we set them to zero.

\paragraph{$N(1520)$:}
In this case, the structure of the  contribution of this resonance to the amputated amplitudes 
 is similar to that of the $\Delta(1232)$, differing just in the definition
  of the appropriate form factors and the isospin dependence. Thus, we have
\begin{eqnarray}
\Gamma^{\mu \rho}_{D_{13}} = \Gamma^{\mu \rho}_{D_{13} P} + \Gamma^{\mu
  \rho}_{CD_{13} P} &=& ie\,
\gamma^0 \left[ J^{\alpha \rho}_{EM (D_{13})}(p',k_\gamma)\right]^\dagger
\gamma^0\frac{P_{\alpha\beta}^{D_{13}}
  (p+q)}{(p+q)^2-M_{D_{13}}^2
+i M_{D_{13}} \Gamma_{D_{13}}} J^{\beta \mu}_{NC (D_{13})}(p,q) \nonumber \\   
&+& ie\, \gamma^0 \left[  J^{\alpha \mu}_{NC (D_{13})}(p^{\,\prime},-q) \right]^\dagger
\gamma^0 \frac{P_{\alpha\beta}^{D_{13}}
  (p^{\,\prime}-q)}{(p^{\,\prime}-q)^2-M_{D_{13}}^2
+i \epsilon} J^{\beta \rho}_{EM (D_{13})}(p,-k_\gamma)
\end{eqnarray}
where the resonance mass $M_{D_{13}}$ is given in Table~\ref{tab:resonances} and the width  $\Gamma_{D_{13}}$ 
is discussed in Appendix~\ref{sec:width};
$P_{\mu\nu}^{D_{13}}$ is the spin 3/2 projection operator given also by 
Eq.~(\ref{eq:spin32}), with the obvious replacement of $M_\Delta$ by $M_{D_{13}}$.
Besides, the $N-N(1520)$ EM and NC transition currents are given by
\begin{eqnarray}
J^{\beta\mu}_{NC(D_{13})}(p,q) &=&  \frac{\tilde{C}^V_{3(D_{13})}(q^2)}{M} (g^{\beta\mu} \slashchar{q}
    -q^\beta \gamma^\mu ) +\frac{\tilde{C}^V_{4(D_{13})}(q^2)}{M^2} (g^{\beta\mu} q \cdot p_{D_{13}}
    -q^\beta p^\mu_{D_{13}} ) +\frac{\tilde{C}^V_{5(D_{13})}(q^2)}{M^2} (g^{\beta\mu} q \cdot p
    -q^\beta p^\mu ) \nonumber \\
&+&\left[ \frac{\tilde{C}^A_{3(D_{13})}(q^2)}{M} (g^{\beta\mu}\slashchar{q}-q^\beta \gamma^\mu ) 
+ \frac{\tilde{C}^A_{4(D_{13})}(q^2)}{M^2} (g^{\beta\mu} q \cdot p_{D_{13}}
    -q^\beta p^\mu_{D_{13}} ) + \tilde{C}^A_{5(D_{13})}(q^2) g^{\beta\mu}\right] \gamma_5
\end{eqnarray}
\begin{eqnarray}
J^{\beta\rho}_{EM(D_{13})}(p,-k_\gamma) &=& -\left[
  \frac{C^V_{3(D_{13})}(0)}{M} (g^{\beta\rho}
  \slashchar{k}_\gamma-k_\gamma^{\beta} \gamma^\rho )
  +\frac{C^V_{4(D_{13})}(0)}{M^2} (g^{\beta\rho} k_\gamma \cdot
  p_{D_{13}\, c} -k_\gamma^\beta p^\rho_{D_{13}\, c} ) \right. \nonumber \\ 
&+& \left. \frac{C^V_{5(D_{13})}(0)}{M^2} (g^{\beta\rho} k_\gamma \cdot p -k_\gamma^\beta p^\rho ) \right],
\end{eqnarray}
where $p_{D_{13}}=p+q$ and $p_{D_{13}\, c}= p-k_\gamma$; 
$\tilde{C}^V_{i(D_{13})}$, $\tilde{C}^A_{i(D_{13})}$ and
$C^V_{i(D_{13})}$ are the NC vector, NC axial and EM form factors,
respectively. The NC vector form factors are related to the EM ones 
in the same way as for the other isospin $1/2$ states considered above, namely 
\begin{eqnarray}
\tilde{C}^{V(p)}_{i(D_{13})}&=&\left(1-4 \sin^2\theta_W
\right) C^{(p)}_{i(D_{13})} -C^{(n)}_{i(D_{13})}
    -C^{(s)}_{i(D_{13})} \,,\nonumber \\
\tilde{C}^{V(n)}_{i(D_{13})}&=& \left( 1-4 \sin^2\theta_W
\right) C^{(n)}_{i(D_{13})}  -C^{(p)}_{i(D_{13})} -C^{(s)}_{i(D_{13})} \,,
\end{eqnarray}
where $C^{(p,n)}_{3-5(D_{13})}$ are obtained from the helicity amplitudes 
using Eqs.~(\ref{eq:a121520}-\ref{eq:s121520}). 
For the axial form factors, one again has that 
\begin{eqnarray}
\tilde{C}^{A(p,n)}_{i(D_{13})} &=& \pm C^A_{i(D_{13})} +
  C^{sA}_{i(D_{13})}  \,, \qquad \left(+\to
p, \, -\to n\right) \,.
\label{eq:c5N1520}
\end{eqnarray}
We take a standard dipole form for the dominant axial NC form factor
\begin{eqnarray}
C^A_{5(D_{13})}(q^2) &=& C^A_{5(D_{13})}(0) \left( 1-\frac{q^2}{M^{* 2}_A} \right)^{-2} \,,
\label{eq:c5q2N1520}
\end{eqnarray}
with $C^A_{5(D_{13})}(0)$ from the corresponding off diagonal GT relation 
(see Appendix~\ref{sec:axialcoupling} and Table~\ref{tab:resonances}), and set 
$M^*_A=1.0$ GeV as for the other $N^*$. The other axial form factors 
$C^A_{3,4(D_{13})}$ are less important because their contribution to the amplitude 
squared is proportional to $q^2$. We neglect them together with the 
unknown strange vector and axial form factors.

\section{Neutral current photon emission in nuclei}
\label{sec:ncgnuclei}
In this section we outline the framework followed
to describe NC photon emission off nuclei. 
Both incoherent and coherent reaction channels are considered.

\subsection{Incoherent neutral current photon emission}
\label{sec:1p1h}
To study the incoherent reactions
\begin{equation}
  \nu_l (k) +\, A_Z   \to \nu_l (k^\prime) + \,
   \gamma(k_\gamma) +\, X,\qquad   \bar\nu_l (k) +\, A_Z   \to \bar\nu_l (k^\prime) + \,
   \gamma(k_\gamma) +\, X,
\label{eq:reacincl}
\end{equation}
we pursue the many body scheme
derived in Refs.~\cite{Nieves:2004wx,Nieves:2005rq,Nieves:2011pp} for 
the neutrino propagation in nuclear matter and adapted to 
(semi)inclusive reactions on finite nuclei by means of 
the local density approximation. With this formalism, the 
photon emission cross section is
\begin{equation}
\left. \sigma_{(\nu,\bar\nu)}\right|_{\mathrm{incoh}} =
    \frac{1}{ |\vec{k}~|}\frac{G^2}{16\pi^2}
     \int  \frac{d^3k'}{|\vec{k}^\prime|}
    L_{\mu\sigma}^{(\nu,\bar\nu)}\left.W^{\mu\sigma}_{{\rm NC}\gamma}\right|_{\mathrm{incoh}} \label{eq:sec-incl}
\end{equation}
in terms of the leptonic tensor of Eq.~(\ref{eq:lep}) and the hadronic tensor $W^{\mu\sigma}_{{\rm NC} \gamma}|_{\mathrm{incoh}}=  W^{(s)\mu\sigma}_{{\rm NC} \gamma}|_{\mathrm{incoh}} + i
W^{(a)\mu\sigma}_{{\rm NC} \gamma}|_{\mathrm{incoh}}$, which is determined by the contributions to the $Z^0$ selfenergy with a photon in the intermediate state  $\Pi^{\mu\sigma}_{Z\gamma}(q)$
\begin{eqnarray}
\left. W^{(s)\mu\sigma}_{NC\gamma}\right|_{incoh} &=& - \Theta(q^0) \left (\frac{4 \cos\theta_W}{g} \right )^2 
\int \frac{d^3 r}{2\pi}~ {\rm Im}\left [ \Pi_{Z\gamma}^{\mu\sigma} 
+ \Pi_{Z\gamma}^{\sigma\mu} \right ] (q,r)\label{eq:zmunuselfs}\\
\left. W^{(a)\mu\sigma}_{NC\gamma}\right|_{incoh}  &=& - \Theta(q^0) \left (\frac{4 \cos\theta_W}{g} \right )^2
 \int \frac{d^3 r}{2\pi}~{\rm Re}\left [ \Pi_{Z\gamma}^{\mu\sigma} 
- \Pi_{Z\gamma}^{\sigma\mu}\right] (q,r) \label{eq:zmunuselfa}.
\end{eqnarray}

\begin{figure}[h!]
\centerline{\includegraphics[height=7.cm]{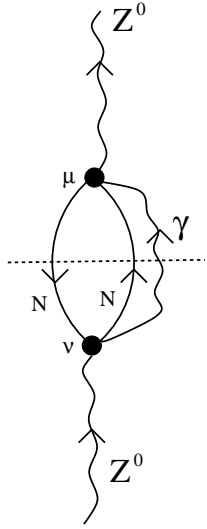}}
\caption{Diagrammatic representation of the one-particle-one-hole-photon (1p1h$\gamma$) 
contributions to the $Z^0$ self-energy in nuclear matter. The black dots represent 
$Z^0N \to \gamma N$ amplitudes.}\label{fig:1p1h}
\end{figure}

In the density expansion proposed in Ref.~\cite{Nieves:2004wx}, the lowest order 
contribution to  $\Pi^{\mu\sigma}_{Z\gamma}$ is  depicted in Fig.~\ref{fig:1p1h}. 
The black dots stand for any of the
eleven terms ($NP$, $CNP$, $\pi Ex$, $RP$, $CRP$ with $R=\Delta(1232)$, $N(1440)$, $N(1520)$, $N(1535)$)
of the elementary $Z^0 N \to \gamma N$  amplitude derived in 
Sec.~\ref{sec:nucleon}. The solid upwards and downwards oriented 
lines represent nucleon particle and hole states in the Fermi sea. 
This $Z^0$ selfenergy diagram (actually 121
diagrams) is readily evaluated as
\footnote{In Eq.~(\ref{eq:W1pibis}), it is necessary to subtract the free space
contribution, i.e., the one that survives for vanishing nuclear
densities and renormalizes free space couplings and masses. 
Actually, to obtain Eq.~(\ref{eq:Z1ga}), we have neglected the
contribution of the antiparticle pole ($p^0=- E(\vec{p}\,)-{\rm
i}\epsilon$) in the $p^0$ integration. This automatically removes 
the unwanted vacuum part.}
\begin{equation}
-i \Pi_{Z\gamma;1{\rm p1h}\gamma}^{\mu\nu}(q,r) = i \left (\frac{g}{4\cos\theta_W} \right )^2 
\sum_{N=p,n} \int \frac{ d^4 k_\gamma}{(2\pi)^4} \int\frac{d^4p}{(2\pi)^4}
\frac{1}{k_\gamma^2+i\epsilon}
{\rm Tr}\left[ S(p,\rho_N)  \gamma^0\left(\Gamma_N^{ \mu \rho}
  \right)^\dagger\gamma^0 S(p',\rho_N) (\Gamma_N)^\nu_{.\, \rho}
\right] \label{eq:W1pibis} 
\end{equation}
where $p'=p+q-k_\gamma$ and $\Gamma_N^{\mu\rho}$ is the amputated amplitude for the 
$Z^0N\rightarrow N \gamma $ process
\begin{gather} 
\Gamma_N^{\mu\rho} = 
\sum_a \Gamma^{\mu \rho}_{a;N},\, 
\quad a=NP,\, CNP,\,\pi Ex,\, RP,\, CRP \,\,
\left [R=\Delta(1232), N(1440), N(1520), N(1535) \right ] \,.
\label{eq:defjA} 
\end{gather}
The nucleon propagator in the medium reads
\begin{equation}
 S(p,\rho_N) = (\slashchar{p}+M) G(p,\rho_N) \,
\end{equation}
with 
\begin{eqnarray}
G(p\,; \rho_N) &=& \frac{1}{p^2-M^2+{\rm
    i}\epsilon}  + {\rm
  i}\frac{\pi}{E(\vec{p}\,)}n_N(\vec{p}\,)\delta(p^0-E(\vec{p}\,))
\label{eq:Gpbis}\\
&=& \frac{1}{p^0+ E(\vec{p}\,) + {\rm
    i}\epsilon} \left ( \frac{n_N(\vec{p}\,)}{p^0- E(\vec{p}\,) - {\rm
    i}\epsilon} + \frac{1-n_N(\vec{p}\,)}{p^0- E(\vec{p}\,) + {\rm
    i}\epsilon}  \right ) \,.  
  \label{eq:Gp}
\end{eqnarray}
The occupation number in the local Fermi gas  
$n_N (\vec{p}\,) = \Theta(k_F^N-|\vec{p}\,|)$ depends on the local density of nucleons (protons or neutrons) 
in the nucleus via $k^N_F(r)= (3\pi^2\rho_N(r))^{1/3}$. The nucleon energy $E(\vec{p})$ is approximated by the 
free one $\sqrt{\vec{p}\,^2 + M^2}$. Substituting the explicit expressions of $S(p,\rho_N)$ and  $S(p',\rho_N)$ in 
Eq.~(\ref{eq:W1pibis}) one obtains
\begin{eqnarray}
-i \Pi_{Z\gamma;1{\rm p1h}\gamma}^{\mu\nu}(q,r)
&=&-\left(\frac{g}{4\cos\theta_W} \right )^2\sum_{N=p,n} \int \frac{
d^4 k_\gamma}{(2\pi)^4}  \int\frac{d^3p}{(2\pi)^3}\frac{1}{2E(\vec{p}\,)} 
\frac{1}{2E(\vec{p}+\vec{q}-\vec{k}_\gamma)}\frac{n_N(\vec{p}\,)[1-n_{N} 
(\vec{p}+\vec{q}-\vec{k}_\gamma)]}{q^{0}-k_\gamma^{0}+E(\vec{p}\,)-E(\vec{p}\,')+i\epsilon}
 \times \nonumber \\
&  \times & \frac{1}{k_\gamma^2+i\epsilon} 
 {\rm
  Tr}\left [(\slashchar{p}+M)\gamma^0\left(\Gamma^{ \mu \rho}_{N}
  \right)^\dagger\gamma^0(\slashchar{p}'+M)\left(\Gamma_{N}\right)^\nu_{.\,\rho}
\right]
  + \left[ (q-k_\gamma)\leftrightarrow -(q-k_\gamma) \right ]\,. \label{eq:Z1ga}
\end{eqnarray}

A convenient simplification can be made by evaluating the
$\Gamma^{\mu\rho}_{N}$ amplitudes at an average nucleon hole
four momentum $\langle p^\mu \rangle$. This 
allows us to take the spin trace in Eq.~(\ref{eq:Z1ga}) out of the $d^3p$ integration, 
which gives, up to constants, the Lindhard function, $\overline
{U}_R(q-k_\gamma,k_F^N,k_F^{N})$ (see Appendix B of
Ref.~\cite{Nieves:2004wx} for definition and explicit expressions). Therefore,  
\begin{eqnarray}
-i \Pi_{Z\gamma;1{\rm p1h}\gamma}^{\mu\nu}(q,r)&=&-\left
(\frac{g}{4\cos\theta_W} \right )^2 \frac{1}{4M^2}\sum_N \int \frac{
  d^4 k_\gamma}{(2\pi)^4}
 \frac{1}{k_\gamma^2+i\epsilon}\overline {U}_R(q-k_\gamma,k_F^N,k_F^{N}) 
A^{\mu\nu}_{N=p,n}\left(\langle p\rangle,q,k_\gamma\right)\,, \label{eq:1ph1ga}
 \\ 
A^{\mu\nu}_{N} &=& \frac12 {\rm
Tr}\left[\left(\langle\slashchar{p}\rangle+M\right)\gamma^0 \left( \left\langle \Gamma_{N}  \right \rangle^{
  \mu \rho} \right)^\dagger \gamma^0\left(\langle\slashchar{p}\rangle+\slashchar{q}
-\slashchar{k}_\gamma+M\right) \left\langle \Gamma_{N}\right \rangle^\nu_{.\,\rho} \right] 
\end{eqnarray}
\noindent
where $\langle \Gamma_{N} \rangle^{\nu\rho}$ stands for $\Gamma^{\nu\rho}_{N}$ calculated at 
the average hole four momentum $\langle p^\mu \rangle$.  

To derive the $1{\rm p1h}\gamma$
contribution to the hadron tensor $W^{\mu\sigma} $,  we remind that by
construction
\begin{equation}
A^{\mu\nu}_{N} = A^{(s) \mu\nu}_{N} +  i
   A^{(a) \mu\nu}_{N} \label{eq:sym}
\end{equation}
where $A^{(s) \mu\sigma}_{N}$
($A^{(a) \mu\sigma}_{N}$) is a real symmetric
(antisymmetric) tensor. Furthermore, it is easy to see that the 
combinations of the $Z^0$ selfenergy present in 
Eqs.~(\ref{eq:zmunuselfs}) and (\ref{eq:zmunuselfa}) fulfill 
\begin{equation}
{\rm Im}\left [ \Pi_{Z\gamma;1{\rm p1h}\gamma}^{\mu\nu} 
+ \Pi_{Z\gamma;1{\rm p1h}\gamma}^{\nu\mu} \right ] = 2  {\rm Im} \Pi_{Z\gamma;1{\rm
    p1h}\gamma}^{(s) \mu\nu}, \qquad   {\rm Re}\left [
  \Pi_{Z\gamma;1{\rm p1h}\gamma }^{\mu\nu} 
- \Pi_{Z\gamma;1{\rm p1h}\gamma}^{\nu\mu} \right ] = -2  {\rm Im} \Pi_{Z\gamma;1{\rm
    p1h}\gamma}^{(a) \mu\nu}
\end{equation}
where $\Pi_{Z\gamma;1{\rm p1h}\gamma}^{(s,a) \mu\nu}$ are obtained by replacing 
$A^{\mu\nu}_{N}$ in Eq.~(\ref{eq:1ph1ga}) by the corresponding 
$A^{(s,a) \mu\nu}_{N}$ parts. 

The imaginary
part of $\Pi_{ZW;1{\rm p1h}\gamma}^{\mu\nu}\Big|_{s(a)}$ can be obtained
following the Cutkosky rules. In this case we
cut the selfenergy diagram of Fig.~\ref{fig:1p1h} with a straight horizontal line. 
The states intercepted by the line are  placed on shell by taking the imaginary part of their
propagators. Technically, the rules to obtain ${\rm Im}
\Pi_{Z;1{\rm p1h}\gamma}^{\mu\nu}$ consist of the following  substitutions:
\begin{eqnarray}
\Pi_{Z\gamma}^{\mu\nu}(q)&\rightarrow& 2i {\rm Im}\Pi^{\mu\nu}_{Z\gamma}(q)\Theta(q^0)\\
\frac{1}{k_\gamma^2+i\epsilon}
 &\rightarrow& 2i {\rm Im} \frac{1}{k_\gamma^2+i\epsilon}
\Theta(k_\gamma^0)  = -2\pi
i \delta(k_\gamma^2) \Theta(k_\gamma^0) \\
\overline {U}_R(q-k_\gamma,k_F^N,k_F^{N}) &\rightarrow& 2i {\rm
  Im}\overline {U}_R(q-k_\gamma,k_F^N,k_F^{N}) \Theta(q^0-k_\gamma^0) \
\end{eqnarray}           
Thus, taking into account that $A^{(s,a) \mu\nu}_{N}$ are real, we readily obtain
\begin{equation}
W^{\mu\nu}_{1{\rm p1h}\gamma}(q) = \Theta(q^0) \frac{1}{2M^2} \int
\frac{d^3r }{2\pi} 
\sum_{N=p,n} \frac{d^3k_\gamma}{(2\pi)^3}
\frac{\Theta(q^0-E_\gamma)}{2E_\gamma} {\rm
  Im}\overline {U}_R(q-k_\gamma,k_F^N,k_F^{N}) A^{\nu\mu}_{N\gamma} \label{eq:1p1hga-def}
\end{equation}
with $E_\gamma$ the photon on-shell energy.  

The average nucleon hole momentum  $\langle p^\mu \rangle$ is chosen as follows  (see the 
discussion after Eq.~(9) of Ref.~\cite{Hernandez:2013jka})
\begin{equation}
\langle p^0 \rangle = \frac{E_F^N + E_\mathrm{min}}2 \,, \qquad \langle
|\vec{p}\,| \rangle = \sqrt{\langle p^0 \rangle^2 -M^2} \label{eq:approx1}
\end{equation}
defined by the central value of the allowed energy region, with 
\begin{equation}
E_\mathrm{min} = {\rm max} \left ( M, E_F^N-q^{\prime 0}, \frac{-q^{\prime
    0}+|\vec{q}^{\,\prime}|\sqrt{1-4M^2/q^{\prime 2}}}2 \right ),
\end{equation}
where $q' =  q-k_\gamma$ and $E_F^N=\sqrt{M^2+(k_F^N)^2}$. The corresponding 
nucleon hole angle, in the LAB frame and with respect to $\vec{q}^{\,\prime}$, 
is completely fixed by the kinematics to 
\begin{equation}
\cos\theta_N = \frac{q^{\prime 2}+2\langle p^0 \rangle q^{\prime 0}}{2 \langle
|\vec{p}\,| \rangle |\vec{q}^{\,\prime}|} 
\label{eq:approx3}
\end{equation}
while the azimuthal angle $\phi_N$ is fixed arbitrarily in the plane 
perpendicular to $\vec{q}^{\,\prime}$. Similar approximations were performed, 
and shown to be sufficiently accurate, in studies
of total inclusive and pion production in photo and electro-nuclear 
reactions~\cite{Carrasco:1989vq,Carrasco:1991mb, Gil:1997bm, Gil:1997jg}. They were also used in 
Ref.~\cite{Nieves:2011pp} to compute the total inclusive neutrino induced cross
section. We have checked that the approximation of Eqs.~(\ref{eq:approx1})--(\ref{eq:approx3}) 
induces uncertainties of at most 5\%, independently of $\phi_N$ values.  
Furthermore, different choices of $\phi_N$ 
produce small variations of the order of 1-2\% in the results.  This 
approximation saves a considerable amount of computational time
because there are analytical expressions for ${\rm Im}\overline
{U}_R(q-k_\gamma,k_F^N,k_F^{N}) $ (see for instance Ref.~\cite{Nieves:2004wx}).

In the small density limit $ {\rm Im}\overline {U}_R(q^{\,\prime},k_F^N,k_F^{N})
\simeq - \pi \rho_N M\delta \left(q^{\,\prime \,0}+M -
\sqrt{M^2+\vec{q}^{\,\prime 2}}\right)/\sqrt{M^2+\vec{q}^{\,\prime 2}}$. Substituting
this expression in Eq.~(\ref{eq:1p1hga-def}) one obtains
\begin{eqnarray}
\lim_{\rho\to 0}W^{\mu\nu}_{1{\rm p1h}\gamma} \sim \int
d\Omega(\hat{k}_\gamma) dE_\gamma
E_\gamma \left (Z W^{\mu\nu}_{Z^0p \to
p\gamma} + N W^{\mu\nu}_{Z^0n \to
n\gamma}\right) \,, 
\end{eqnarray}
where $Z$ and $N$ are the number of protons and neutrons in the nucleus,
and $W^{\mu\nu}_{Z^0N \to N\gamma}$ is the hadronic tensor for NC
photon production on the nucleon. In this way, the strict impulse
approximation is recovered. By performing the integral in
Eq.~(\ref{eq:1p1hga-def}), Pauli blocking and Fermi motion are taken into account.

\subsubsection{Further nuclear medium corrections}
\label{sec:delta-medium}
Given the dominant role played by the $\Delta P$ contribution and
since $\Delta$ properties are strongly modified in the nuclear
medium~\cite{Hirata:1978wp,Oset:1981ih,Freedman:1982yp,Oset:1987re,Nieves:1993ev,
Singh:1998ha,Lehr:1999zr} a proper treatment of the
$\Delta$ contribution is needed. Here, we follow
Ref.~\cite{AlvarezRuso:2007tt} and modify the $\Delta$ propagator in the
$\Delta P$ term as 
\begin{equation}
\frac1{p_\Delta^2-M_\Delta^2+iM_\Delta\Gamma_\Delta}\to\frac1{\sqrt{p_\Delta^2}+M_\Delta}
\frac1{\sqrt{p_\Delta^2}-M_\Delta+i(\Gamma^{\rm Pauli}_\Delta/2-{\rm
    Im}\Sigma_\Delta)}\,; 
\end{equation}
$\Gamma^{\rm Pauli}_\Delta$, for which we take the expression in Eq. (15) of
Ref.~\cite{Nieves:1991ye},  
is the free $\Delta$ width corrected by the Pauli blocking
of the final nucleon. The imaginary
part of the $\Delta$ self-energy in the medium ${\rm Im}\Sigma_\Delta$,  
is parametrized as~\cite{Oset:1987re} 
\begin{equation}
-\mathrm{Im}\Sigma_\Delta(\rho)=C_Q\left(\frac{\rho}{\rho_0}\right)^{\alpha}+C_{A2}\left(\frac{\rho}{\rho_0}\right)^{\beta}+
C_{A3}\left(\frac{\rho}{\rho_0}\right)^{\gamma}\,, \label{eq:delta-self}
\end{equation}
where the term proportional to $C_Q$ accounts for the QE part while 
those with coefficients $C_{A2}$ and $C_{A3}$ correspond to the two-body ($\Delta
N\rightarrow NN$) and three-body ($\Delta NN\rightarrow NNN$)
absorption contributions, respectively. The parameters in Eq.~(\ref{eq:delta-self}) 
 can be found in Eq.~(4.5) and Table 2 of Ref.~\cite{Oset:1987re}, given as 
functions of the kinetic energy in the laboratory system of a pion that
would excite a $\Delta$ with the corresponding invariant mass.
These parametrizations are valid in the range 85 MeV
$< T_\pi<$315 MeV. Below
85 MeV, the contributions from $C_Q$ and $C_{A3}$ are rather small
and are taken from Ref~\cite{Nieves:1991ye}, where the model was extended
to low energies. The term with $C_{A2}$ shows a very mild energy
dependence and we still use the parametrization from Ref.~\cite{Oset:1987re} 
even at low energies. For $T_\pi$ above 315 MeV we have kept
these self-energy terms constant and equal to their values at the
bound. The uncertainties in these pieces are not very relevant
there because the $\Delta \rightarrow N\pi$ decay becomes very large and
absolutely dominant.
  
For the $\Delta$ mass we shall keep its free value. While there are some 
corrections arising from both the real part of the self-energy and random
phase approximation (RPA) sums, the net effect is smaller than the precision 
achievable in current neutrino experiments,
and also smaller than the uncertainties due to our limited knowledge of the
nucleon to $\Delta$ transition form factor $C_5^A(q^2)$ (see
the related discussion in Sec. II.E of Ref.~\cite{Nieves:2011pp}).

\subsection{Coherent neutral current photon emission}
\label{sec:coherent}

The coherent reactions 
\begin{equation}
  \nu_l (k) +\, A_Z|_{gs}(p_A)  \to \nu_l (k^\prime) +
  A_Z|_{gs}(p^\prime_A) +\, \gamma(k_\gamma), \qquad   
\bar\nu_l (k) +\, A_Z|_{gs}(p_A)  \to \bar\nu_l (k^\prime) +
  A_Z|_{gs}(p^\prime_A) +\, \gamma(k_\gamma) 
\label{eq:reac}
\end{equation}
consist of a weak photon production where the nucleus
is left in its ground state, in contrast with the incoherent
production that we studied in the previous subsection, where the
nucleus is either broken or left in an excited state. Here, we 
adopt the framework derived in Ref.~\cite{Amaro:2008hd} for neutrino-induced
coherent CC and NC pion production reactions.\footnote{The predictions of
  Ref.~\cite{Amaro:2008hd} were updated in \cite{Hernandez:2010jf}
  after the reanalysis of the $\nu_\mu p \to \mu^- p \pi^+$ old bubble
  chamber data carried out in Ref.~\cite{Hernandez:2010bx}.} This work is,
in turn, based on previous studies of coherent pion production in
electromagnetic [$(\gamma, \pi^0)$~\cite{Carrasco:1991we}, $(e,
  e'\pi^0)$~\cite{Hirenzaki:1993jc}] and hadronic reactions [$(^3{\rm
    He}, ^3{\rm H}\, \pi^+)$~\cite{FernandezdeCordoba:1992ky}, 
$p(^4{\rm He},^4{\rm He})X$~\cite{FernandezdeCordoba:1993az}] in the
$\Delta (1232)$ region.  More recently, the same scheme has been employed to
study charged kaon production by coherent scattering of neutrinos and
antineutrinos on nuclei~\cite{AlvarezRuso:2012fc}.  The model for the
coherent process is built up from the coherent scattering 
with each of the nucleons of the nucleus, producing an
outgoing $\gamma$. The nucleon state (wave function) remains unchanged 
so that after summing over all nucleons, one
obtains the nuclear densities. In the elementary $Z^0 N \to N \gamma$
process, energy conservation is accomplished by imposing
$q^0=E_\gamma$, which is justified by the large nucleus mass, 
while the transferred momentum
$\vec{q}-\vec{k}_\gamma$ has to be accommodated by the nucleon wave
functions. Therefore, the coherent production process is sensitive to the
Fourier transform of the nuclear density.

Following Ref.~\cite{Amaro:2008hd}, it is straightforward to find that
\begin{gather}
\left.\frac{d^{\,3}\sigma_{(\nu,\bar\nu)}}{dE_\gamma
  d\Omega(\hat{k}_\gamma)}\right|_{\rm coh}  =
    \frac{E_\gamma}{ |\vec{k}~|}\frac{G^2}{16\pi^2}
     \int  \frac{d^3k'}{|\vec{k}^\prime|}
    L_{\mu\sigma}^{(\nu,\bar\nu)}\left.W^{\mu\sigma}_{{\rm
        NC}\gamma}\right|_{\rm coh} \,, \\[0.2cm]
 \left. { W}^{\mu\sigma}_{{\rm NC}\gamma}\right|_{\rm coh} = -
  \frac{\delta(E_\gamma-q^0)}{64\pi^3M^2}  {\cal A}^{\mu\rho}(q,k_\gamma)
    \left({\cal A}^\sigma_{.\,\rho}\right)^*(q,k_\gamma) \label{eq:zmunu} \,,
\\[0.2cm] 
{\cal A}^{\mu\rho}(q,k_\gamma) = 
\int d^3r\ e^{i\left(\vec{q}-\vec{k}_\gamma\right)\cdot\vec{r}} 
\left\{\rho_{p}(r\,) {\hat \Gamma}^{\mu\rho}_{p}(r;q,k_\gamma) 
+ \rho_{n}(r\,)
    {\hat \Gamma}^\mu_{n}(r;q,k_\gamma) \right\}\label{eq:Jmunu2} \,.
    \end{gather}
To evaluate the hadronic tensor, we use the model for the NC photon
production off the nucleon derived in Sect.~\ref{sec:nucleon} and thus
we have
\begin{gather} 
{\hat \Gamma}^{\mu\rho}_{N}(r;q,k_\gamma) = 
\sum_i {\hat \Gamma}^{\mu\rho}_{i;N}(r;q,k_\gamma),\, 
\quad i=NP,\, CNP,\,\pi Ex,\, RP,\, CRP \,\,
\left [R=\Delta,N(1440), N(1535), N(1520)\right ]
\label{eq:cn-jcoh} \\
  \left. {\hat \Gamma}^{\mu\rho}_{i;N}(r;q,k_\gamma) = 
\frac12 {\rm Tr}\left[(\slashchar{p}+M)\gamma^0\,\Gamma_{i;
  N}^{\mu\rho} \right]\frac{M}{p^0}  
  \right|_{p^\mu=\left (\sqrt{M^2+\frac{(\vec{k}_\gamma-\vec{q}\,)^2}{4}},
      \frac12(\vec{k}_\gamma-\vec{q}\,)\right )} 
\label{eq:cn-jcoh2}
\end{gather}
where the four-vector matrices $\Gamma_{i; N\gamma}^{\mu\rho}$ stand
for the amputated photon production amplitudes off
nucleons derived in Subsec.~\ref{sec:gamma_amp}.
We have also taken into account the modification of the $\Delta(1232)$ 
in the medium for the $\Delta P$ mechanism, as explained in
Subsec.~\ref{sec:delta-medium}. 

Now we pay attention to the approximated treatment of nucleon momentum 
distributions that has been adopted to obtain
Eqs.~(\ref{eq:zmunu})--(\ref{eq:cn-jcoh2}). The initial ($\vec{p}$)
and final ($\vec{p}^{\,\prime}$) nucleon three momenta are not
well defined. We take
\begin{equation}
p^\mu = \bigg( \sqrt{M^2+ \frac14{\left(\vec{k}_\gamma-\vec{q}\right)^2}} ,
  \frac{\vec{k}_\gamma-\vec{q}}{2}\,\bigg)  \label{eq:pmu} \, , \qquad
p^{\prime\,\mu} = q - k_\gamma + p
= \bigg( \sqrt{M^2+ \frac14{\left(\vec{k}_\gamma-\vec{q}\right)^2}} ,
  -\frac{\vec{k}_\gamma-\vec{q}}{2}\,\bigg)\,,
\end{equation}
with both nucleons being on-shell.  In this way, the
momentum transfer is equally shared between the initial and final
nucleons. This prescription, employed in 
Refs.~\cite{AlvarezRuso:2007tt,AlvarezRuso:2007it,Amaro:2008hd,Zhang:2012xi},
for (anti)neutrino induced coherent pion production, was earlier applied to 
$^{16}\mathrm{O}(\gamma,\pi^+)^{16}\mathrm{N}_\mathrm{bound}$~\cite{Eramzhian:1983pe} 
and to coherent $\pi^0$ photo- and 
electroproduction~\cite{Carrasco:1991we,Hirenzaki:1993jc,Drechsel:1999vh}. 
The approximation is based on the fact that, for Gaussian nuclear wave
functions, it leads to an exact treatment of the terms in the elementary amplitude 
that are linear in
momentum. In Ref.~\cite{Carrasco:1991we}
it was shown that in the case of $\pi^0$ photoproduction,  
this prescription provided similar results as the
explicit sum over the nucleon momenta performed in
Ref.~\cite{Boffi:1991nh}. Thanks to the choice of Eq.~(\ref{eq:pmu}),  
the sum over all nucleons is greatly simplified and cast in terms of
the neutron and proton densities [see Eq.~(\ref{eq:Jmunu2})]. 
Furthermore, the sum over nucleon helicities gives rise to the trace in
Eq.~(\ref{eq:cn-jcoh2}); more details can be found in the discussion after 
Eq.~(6) of Ref.~\cite{Amaro:2008hd}. On the other hand, 
 this approximation eliminates some non-local
contributions to the amplitudes. In particular, the $\Delta$ momentum 
turns out to be well defined once the the nucleon momenta are fixed.
In Ref.~\cite{Leitner:2009ph} this
constraint was relaxed for weak coherent pion
production via $\Delta(1232)$ excitation,  while neglecting the modification of the  
$\Delta$ properties in the nucleus and pion distortion. 
It was found that non-localities in the $\Delta$
propagation cause a large reduction of the cross section at low
energies. In the more realistic description of Nakamura et
al.~\cite{Nakamura:2009iq}, the non-locality is preserved for the 
$\Delta$ kinetic term in a linearized version of the $\Delta$ propagator 
but, at the same time, a prescription similar to Eq.~(\ref{eq:pmu}) 
for the $WN\Delta$ and $\Delta N\pi$ vertices, and a local 
ansatz for the in-medium $\Delta$ selfenergy have been taken. Nevertheless, 
the mismatch between the non-local recoil effects and the local approximations 
for vertices and selfenergy are likely to be minimized by the fact that the 
parameters in the $\Delta$ selfenergy are adjusted to describe pion-nucleus 
scattering data with the same model. Our point of view is that  
the local approach adopted here and in 
Refs.~\cite{AlvarezRuso:2007tt,AlvarezRuso:2007it,Amaro:2008hd,Zhang:2012xi}, 
together with the choice of the  effective nucleon-nucleon interaction in the medium~\cite{Oset:1987re}, 
is internally consistent. The good agreement obtained for pion-nucleus 
scattering~\cite{GarciaRecio:1989xa,Nieves:1991ye} and coherent pion photoproduction~\cite{Zhang:2012xi,JuanPC}
for medium and heavy nuclei seems to support this conjecture, although more detailed investigations are 
necessary. In any case, for the present study, where the coherent contribution 
is a small and not disentangled  part of the total $NC\gamma$ cross section, 
and in view of the uncertainty in the determination of the 
$N\Delta$ axial coupling $C^A_5(0)$, it is safe to disregard possible non-local 
corrections.

\section{Results}
\label{sec:results}

Before discussing our results an important remark is due. 
The intermediate nucleon propagators in both the $NP$ and $CNP$ terms
 of Eq.~(\ref{eq:NP-CNP}) can be put on the mass shell for $E_\gamma\to
0$ photons, leading to an infrared divergence. This divergence should be cancelled
by others present in the electromagnetic radiative corrections to
the elastic process $\nu N \to \nu N$ (without photon
emission). However, when the emitted photon is too soft, its energy
becomes smaller than the photon energy resolution of the detector. 
Such an event would be recorded as an elastic one if at all.
For this reason, we have
implemented a cut in the available photon phase space, demanding 
 $E_\gamma \geq 140$ MeV, which corresponds to the MiniBooNE 
detection threshold~\cite{AguilarArevalo:2007it}. 

\subsection{Neutral current photon emission off nucleons}
\label{sec:res1}

\begin{figure}[h!]
\begin{center}
\makebox[0pt] {\includegraphics[width=0.5\textwidth]{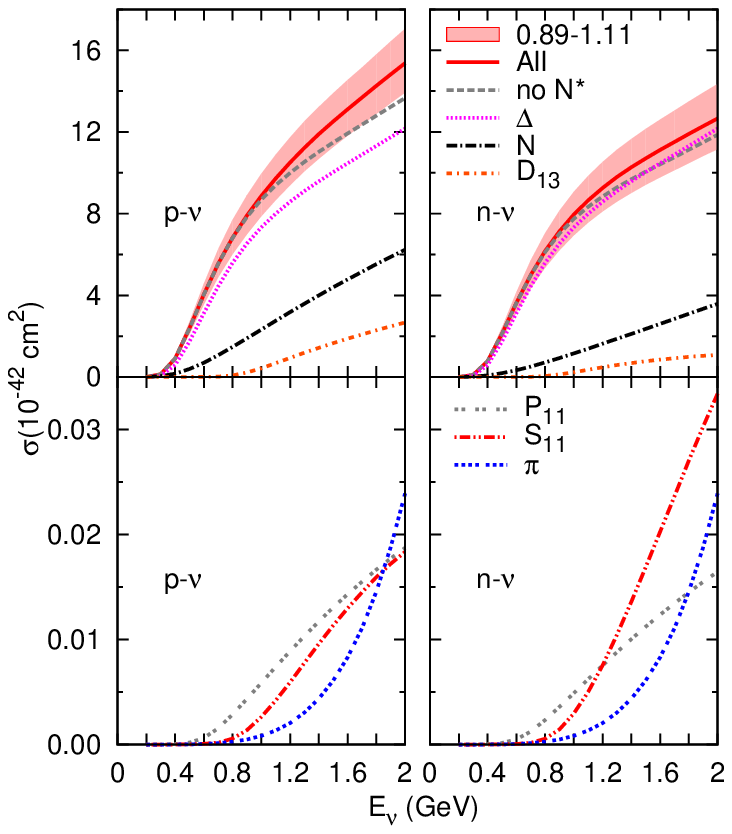}\includegraphics[width=0.5\textwidth]{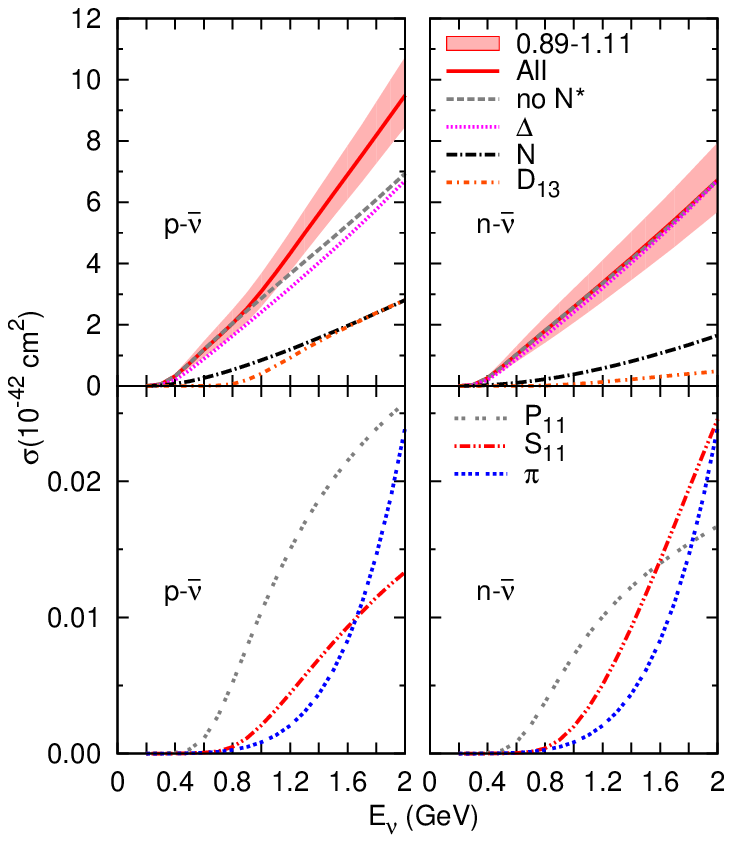}}
\caption{(color online) $\nu N \to \nu N \gamma$ (left) and $\bar\nu N
  \to \bar\nu N \gamma$ (right) cross sections on protons and neutrons as a function of the
  (anti)neutrino energy. A cut of $E_\gamma \geq 140$~MeV 
in the phase space integrals has been applied.
Solid curves correspond to
  the results from the full model, with error bands determined by
  the uncertainty in the  axial $N\Delta$ coupling $C^A_5(0)=1.0 \pm 0.11$ 
according to the determination of Ref.~\cite{Hernandez:2010bx}. The curves labeled as $\Delta$,
  $N$ and $\pi$ stand for the partial contributions of the $(\Delta
  P+C\Delta P)$, $(N P+CNP)$ and $\pi Ex$ mechanisms,
  respectively. The  $D_{13}$, $P_{11}$ and $S_{11}$ curves show the
  contribution of the different $(RP+CRP)$ terms driven by the $N^*$ resonances. 
  Finally, the lines labeled as ``no $N^*$''
  display the predicted cross section without the $N^*$ contributions. }
\label{fig:cs_nu}
\end{center}
\end{figure}
In Fig.~\ref{fig:cs_nu}, we show our results for the total NC photon
emission (anti)neutrino cross sections as a function of the
(anti)neutrino energy. As in other weak interaction processes, 
the different helicities of $\nu$ and
$\bar \nu$ are responsible for different interference patterns, resulting in
smaller $\bar \nu$ cross sections with a more linear energy
dependence. The error bands on the full model results are 
determined by the uncertainty in the
axial $N\Delta$ coupling $C^A_5(0)=1.00\pm0.11$~\cite{Hernandez:2010bx}. 
This is the predominant source of uncertainty in the (anti)neutrino energy 
range under consideration (see also the discussion of Fig.~\ref{fig:delta_term} below).  
We also display the contributions from the different mechanisms considered in our model
(Figs.~\ref{fig:five} and \ref{fig:highR}).  The $\Delta$ mechanism is
dominant and gives the same contribution for protons and neutrons, as 
expected from the isovector nature of the electroweak $N-\Delta$ transition. 
At $E_{\nu(\bar \nu)} \sim 1.5$ GeV, the cross section from nucleon-pole
terms is only about 2.5 smaller than the $\Delta$ one.  Above $\sim 1.5$~GeV,
the $N(1520)$ contribution is sizable and comparable to that of the
sum of the $NP$ and $CNP$ mechanisms, specially for
$\bar \nu p$. However, the rest of $N^*$ contributions considered in the
model (with $N(1440)$ and $N(1535)$ intermediate states), together
with the $\pi Ex$ contribution of Fig.~\ref{fig:five}(e) can be safely
neglected in the whole range of (anti)neutrino energies considered in
this work. The fact that the  $N(1520)$ resonance is the only one, 
besides the $\Delta(1232)$, playing a significant role for $E_\nu < 2$~GeV 
has also been observed in pion production~\cite{Hernandez:2013jka} and 
for the inclusive cross section~\cite{Leitner:2008ue}.

\begin{figure}[h!]
\begin{center}
\makebox[0pt] {\includegraphics[width=0.5\textwidth]{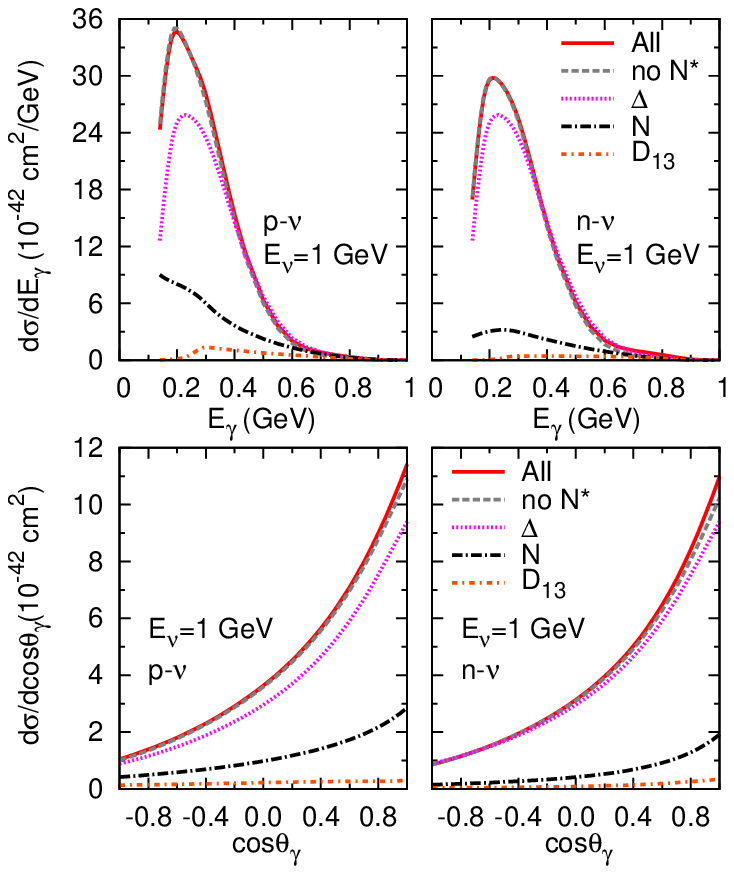}\includegraphics[width=0.5\textwidth]{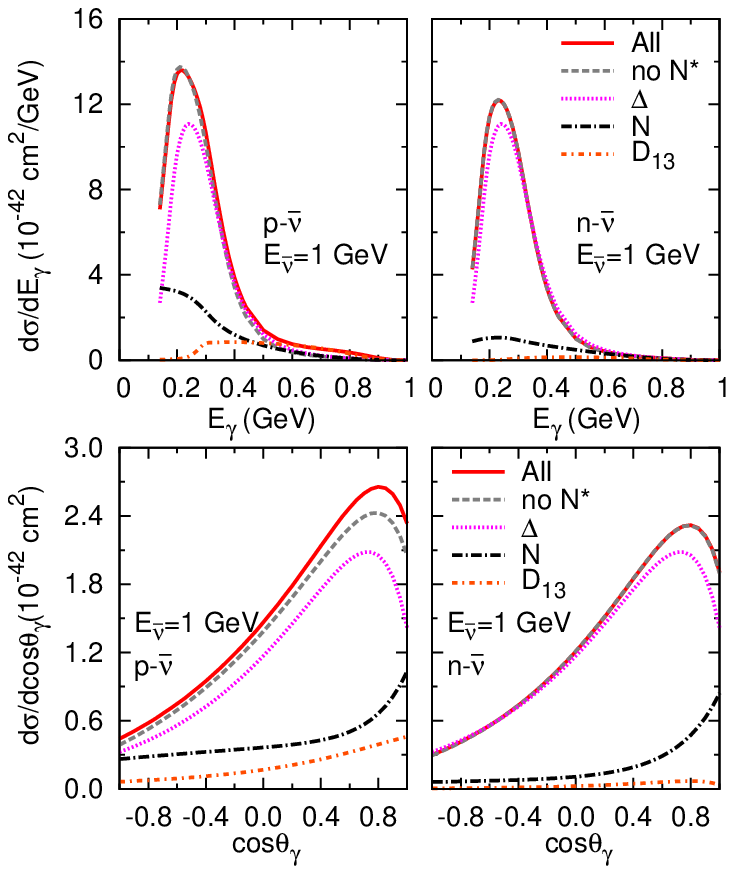}}
\caption{(color online) $\nu N \to \nu N \gamma$ (left) and $\bar\nu N
  \to \bar\nu N \gamma$ (right) photon energy (top) and photon angular
  (bottom) differential cross sections at $E_{\nu,\bar\nu}=1$ GeV on both 
protons and neutrons. The angle
  $\theta_\gamma$ is referred to the direction of the incoming
  (anti)neutrino beam. A cut of $E_\gamma \geq 140$~MeV has been applied.
Solid curves are for the full model. 
 The curves labeled as $\Delta$,  $N$
  and $D_{13}$ stand for the partial contributions of the $(\Delta
  P+C\Delta P)$, $(N P+CNP)$ and the $(N(1520)P+CN(1520)P)$ terms,
  respectively. The lines labeled as ``no $N^*$'' display the
  predictions neglecting the $N^*$ contributions. }
\label{fig:1gev}
\end{center}
\end{figure}
\begin{figure}[h!]
\begin{center}
\makebox[0pt] {\includegraphics[width=0.5\textwidth]{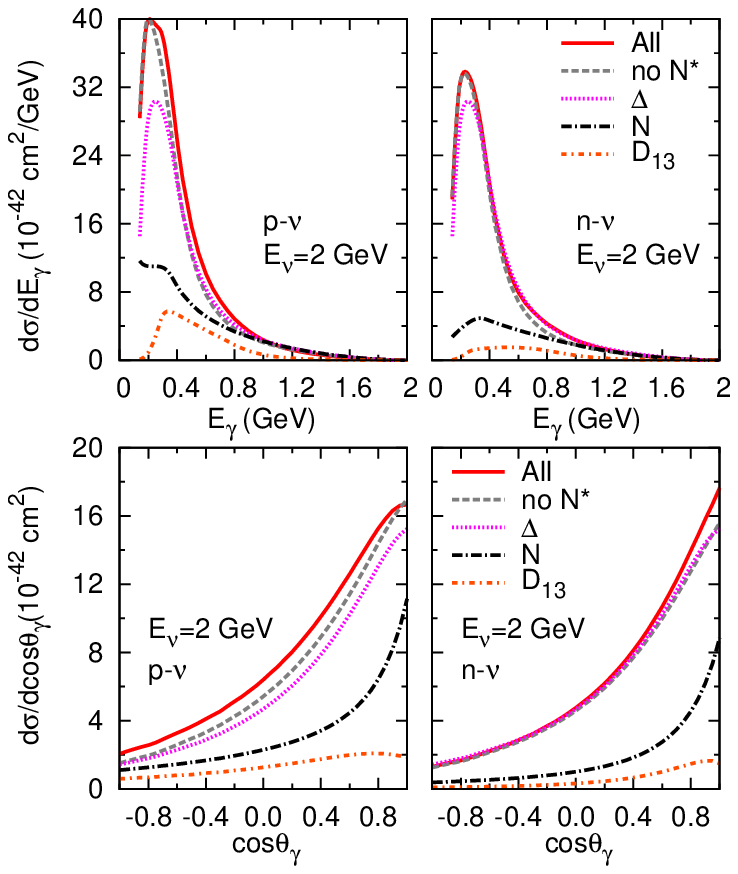}\includegraphics[width=0.5\textwidth]{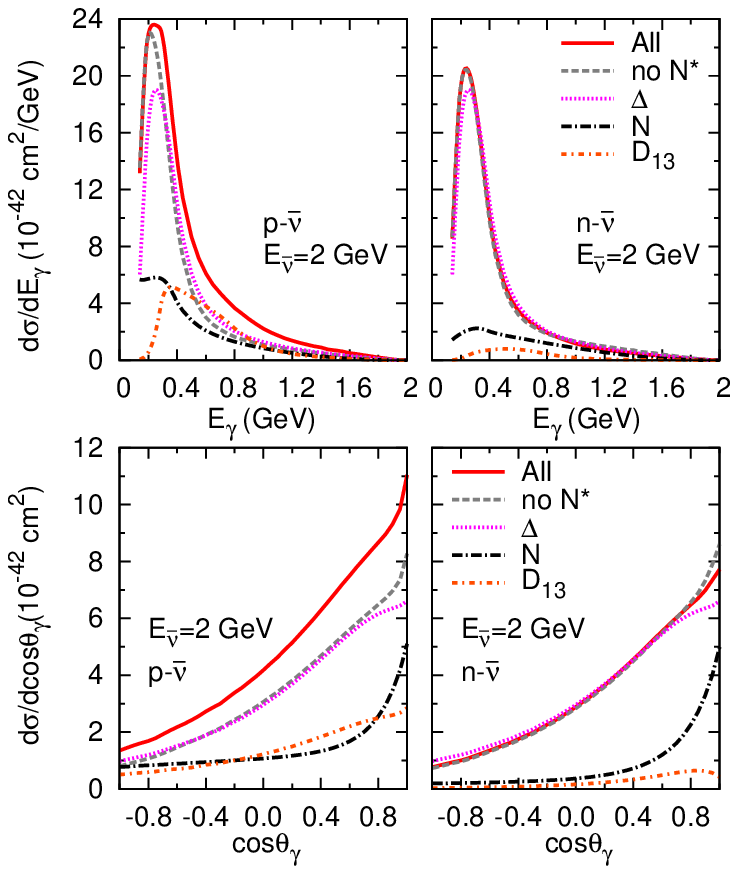}}
\caption{(color online) Same as Fig.~\ref{fig:1gev}, but for an
  (anti)neutrino energy of 2 GeV.}
\label{fig:2gev}
\end{center}
\end{figure}

Photon angular and energy distributions on single nucleons, for
incoming (anti)neutrino energies of 1 and 2 GeV are shown in
Figs.~\ref{fig:1gev} and \ref{fig:2gev}.  Solid curves
stand for the results from the full model. We also display the largest
contributions among the different mechanisms considered in our model.  
As expected, the
$\Delta$ mechanisms are also dominant in the differential
cross sections, specially for reactions on neutrons and even more so 
for the $\bar\nu n \to \bar\nu n \gamma$ process. Nucleon and $D_{13}$
direct and crossed pole-term contributions, though small, are not
negligible, particularly for protons. The $N(1520)$ terms become 
more important for the largest (anti)neutrino energy. 
At the lower energy the reaction is more forward-peaked for neutrinos 
than for antineutrinos. In the later case, the maximum of the distribution 
moves forward as the energy increases.
The photon energy differential cross sections always exhibit 
a peak slightly above $E_\gamma=0.2$ GeV, mainly produced
by the interplay between the $\Delta-$ pole and the three-body phase
space photon energy distribution. The $\Delta$ propagator suppresses 
not only the low photon energy contributions, but also the high photon
energy tail that would appear because of the boost to the LAB
frame. 

Next, we compare our predictions for the nucleon cross sections
with those obtained in Refs.~\cite{Hill:2009ek,Zhang:2012xn}.  These two
models include the $NP+CNP$ and $\Delta P+C\Delta P$ mechanisms, 
with dominance of $\Delta P$ like in our case.  
The Compton-like contributions ($NP+CNP$) are determined by the
electromagnetic and axial nucleon form factors, which are reasonably
well constrained.  The predictions of Ref.~\cite{Zhang:2012xn} for
these mechanisms are similar to ours. Instead, those in
Ref.~\cite{Hill:2009ek} exhibit a steeper energy dependence, because
of the higher nucleon axial mass, $M_A=1.2 $ GeV in $F_A$
[Eq.~(\ref{eq:gaxial})], used there. This choice was motivated by the
first phenomenological analysis of the MiniBooNE CCQE scattering data on 
carbon using the relativistic Fermi gas model~\cite{AguilarArevalo:2007ab}\footnote{In the final MiniBooNE
  analysis~\cite{AguilarArevalo:2010zc}, an even larger value of
  $M_A\sim 1.35$ GeV was obtained. }.  Later theoretical 
studies~\cite{Martini:2009uj, Martini:2011wp,Nieves:2011pp,Nieves:2011yp} 
have shown that such high values of $M_A$ encoded multi-nucleon contributions 
that were not taken into account in the experimental analyses.
We use a lower value for $M_A = 1$~GeV, which is consistent with two 
independent experimental
sources: bubble chamber neutrino/antineutrino induced QE reactions on 
hydrogen and deuterium and pion electroproduction~\cite{Bodek:2007ym}. 
In addition to the $NP+CNP$ and $\Delta P+C\Delta P$ mechanisms,
R. Hill~\cite{Hill:2009ek} also considers $t-$channel $\pi$, $\rho$
and $\omega$ exchanges. Only the latter one provides a non-negligible
cross section that, for antineutrinos, could become
comparable to the nucleon Compton-like contribution for incident 
energies above 1.5 GeV. However, the size of the $\omega$ contribution
strongly depends on the mostly undetermined off-shell form factor and 
is then affected by large uncertainties.

In the model of X. Zhang and B. Serot~\cite{Zhang:2012xn}, additional 
contact terms allowed by symmetry were considered. 
As pointed out in the Introduction, they
notably increase the cross section above $\sim 1$ GeV (see Fig. 3 of
that reference). In Ref.~\cite{Serot:2012rd}, it is argued that these
contact terms are the low-energy manifestation of anomalous $\rho$ and
$\omega$ interactions; their contributions below 550 MeV are
very small, as expected on the base of the power counting established there. 
To extend these findings to higher energies, phenomenological form factors 
are employed~\cite{Zhang:2012xn}, which are, however, not well understood. 
Therefore, their cross section above $E_\nu \sim 1$ GeV should be taken cautiously 
once contact terms are a source of uncontrolled systematics.

We now focus on the comparison for the dominant $\Delta$ contribution, which is 
presented in Fig.~\ref{fig:delta_term}. 
Different values of the axial $N\Delta$ coupling $C^A_5(0)$ and photon
energy cuts have been implemented in Refs.~\cite{Zhang:2012xn,Hill:2009ek}, 
as specified in the caption of
Fig.~\ref{fig:delta_term}. We have used these inputs and compared our
predictions with those found in these references, finding a good 
agreement particularly with Ref.~\cite{Zhang:2012xn}. 
In the case of Ref.~\cite{Hill:2009ek} the
agreement is better for antineutrinos than for neutrinos. 
However, in the actual calculations, a 
major difference arises from the fact that we are using a
substantially lower value of $C^A_5(0)=1.00$. Thus, our final
predictions for the dominant $\Delta$ contribution are about 30\%
or 45\% smaller than those of Refs.~\cite{Zhang:2012xn} and
Ref.~\cite{Hill:2009ek}, respectively. The error bands in our results 
of Fig.~\ref{fig:cs_nu}, which are determined by the uncertainty 
in $C^A_5(0)$,  partially englobe these discrepancies. In this context, 
it is worth reminding that the value of $C^A_5(0)=1.00 \pm 0.11$ used here  
was determined in a
combined analysis of the neutrino induced pion production
ANL~\cite{Barish:1978pj,Radecky:1981fn} and
BNL~\cite{Kitagaki:1986ct,Kitagaki:1990vs} bubble chamber data. This was done 
with a model closely resembling the present one i.e. including nonresonant 
mechanisms, with the correct threshold behavior dictated by chiral symmetry, 
 the dominant $\Delta(1232)$ excitation and also deuteron effects~\cite{Hernandez:2010bx}. 
Such a consistency with pion production data on the nucleon was not attempted in 
Refs.~\cite{Hill:2009ek,Zhang:2012xn}. Actually, the ANL $\nu_\mu p \to \mu^- p \pi^+$ data are notably
overestimated in Ref.~\cite{Zhang:2012xn} as can be seen in Fig. 2 of that article. 
\begin{figure}[h!]
\begin{center}
\includegraphics[width=0.5\textwidth]{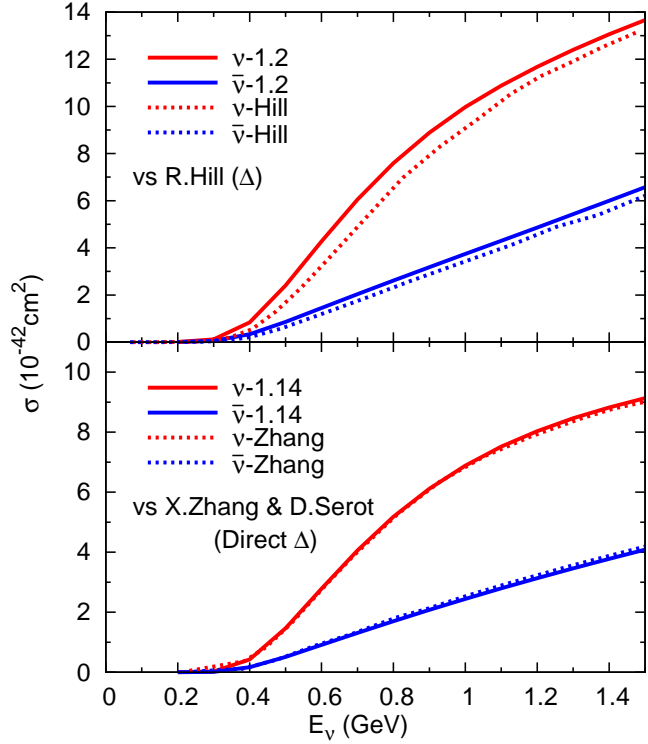} 
\caption{(color online) Top panel: 
 $\Delta P+C\Delta P$ cross sections obtained by us (solid lines) 
and from Ref.~\cite{Hill:2009ek} (dashed lines), for  $\nu N \to \nu N \gamma$ 
(red upper curves) and for $\bar\nu N \to \bar\nu N \gamma$ (blue lower 
curves). For this comparison we have taken  $C^A_5(0) = 1.2$ and no cut in $E_\gamma$
as in Ref.~\cite{Hill:2009ek}. Bottom panel: $\Delta P$ cross section obtained by us (solid lines) 
and from Ref.~\cite{Zhang:2012xn} (dashed lines), for  $\nu N \to \nu N \gamma$ 
(red upper curves) and for $\bar\nu N \to \bar\nu N \gamma$ (blue lower 
curves). For this comparison we have adopted $C^A_5(0) = 1.14$ and an $E_\gamma \geq 0.2$ GeV cut, 
as in Ref.~\cite{Zhang:2012xn}.}
\label{fig:delta_term}
\end{center}
\end{figure}

\subsection{Neutral current photon emission in nuclei}
\label{sec:res2}

For the present computations we take nuclear charge density distributions, 
normalized to the number of protons in the nucleus, extracted from 
electron scattering data~\cite{DeJager:1974dg}. The neutron matter density 
profiles are parametrized in the same way as the charge densities 
(but normalized to the number of neutrons) with small changes from 
Hartree-Fock calculations~\cite{Negele:1972zp} and supported by pionic atom
data~\cite{GarciaRecio:1991wk}. The corresponding parameters are 
compiled in Table I of Ref.~\cite{Nieves:1993ev}. Furthermore, these density 
distributions have been deconvoluted to get center-point densities following the
procedure described in Ref.~\cite{Oset:1989ey}.

\subsubsection{Incoherent reaction: $1p1h\gamma$ contribution}
\label{sec:res2a}

\begin{figure}[h!]
\begin{center}
\makebox[0pt] {\includegraphics[width=0.5\textwidth]{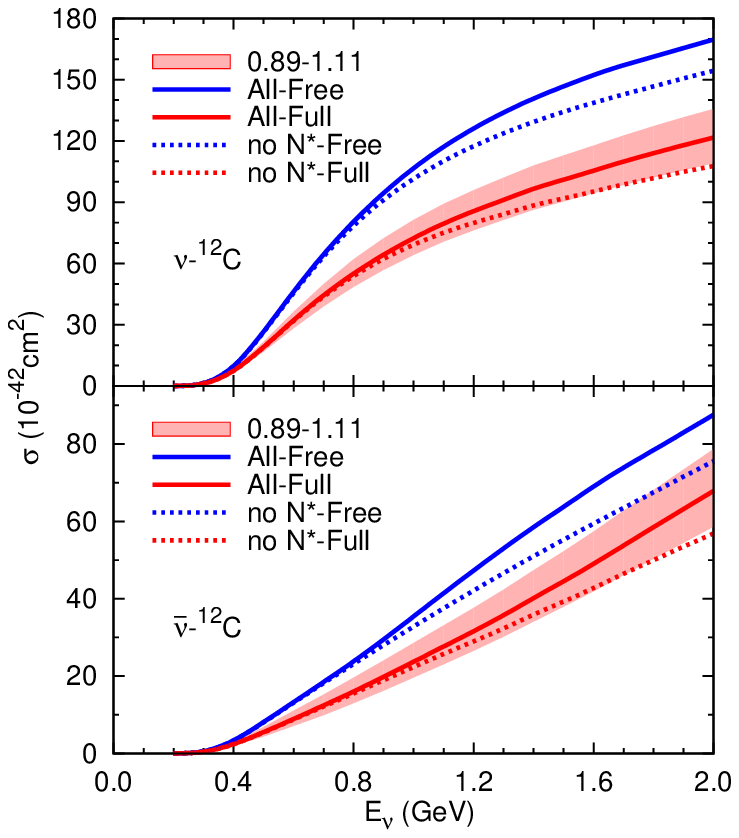}
\includegraphics[width=0.5\textwidth]{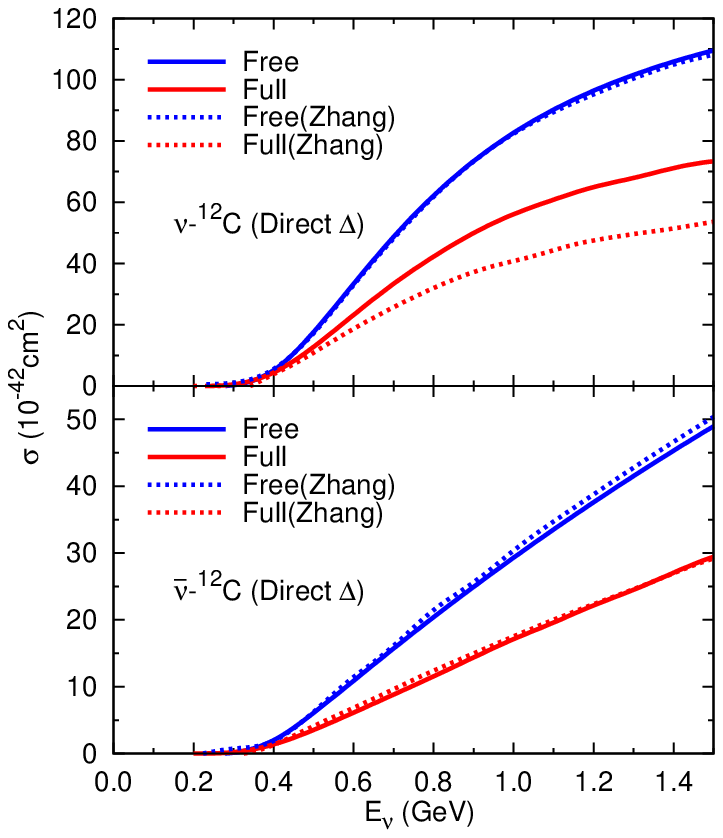}}
\caption{(color online) Left panel: Neutrino (top) and antineutrino (bottom)
   incoherent photon emission cross sections on $^{12}$C.  
All curves in this panel have been obtained with an $E_\gamma \geq 140$~MeV cut in the 
phase space. Solid lines stand for
  results from the complete model at the nucleon level,  
while the dotted lines display
  the predicted cross sections without the $N^*$ contributions.  
  Curves denoted as ``Free'' (upper blue curves) do
  not include any nuclear correction:
  the nuclear target is treated as a mere ensemble of
  nucleons ($\sigma_A=Z\sigma_p+N\sigma_n)$. 
 Curves labeled as ``Full'' (lower red curves) take into account Pauli
  blocking, Fermi motion and the in medium $\Delta$ resonance
  broadening. The error bands show the uncertainty on the full model 
 that arises from the 
 determination of the axial $N\Delta$ coupling from data 
($C^A_5(0) = 1.00 \pm 0.11$)~\cite{Hernandez:2010bx}. 
 Right panel: $\Delta P$ contribution to the neutrino
  (top) and antineutrino (bottom) photon emission cross sections on 
  $^{12}$C from Ref.~\cite{Zhang:2012xn} compared to our predictions 
for the same mechanism, adopting the same infrared photon energy cut $E_\gamma \geq 0.2$~GeV 
and  $C^A_5(0)=1.14$. The meaning of ``Free'' and ``Full'' labels is
  the same as in the left plots.}
\label{fig:incoh1}
\end{center}
\end{figure}

In the left panels of Fig.~\ref{fig:incoh1}, we show our predictions
for the (anti)neutrino incoherent photon emission cross
sections on $^{12}$C as a function of the (anti)neutrino energy up to 2 GeV. 
We observe that the neglect of nuclear medium corrections, as it
was done in the study of the NC$\gamma$ excess of events at MiniBooNE  
of Ref.~\cite{Hill:2010zy}, is a quite poor approximation.  
By taking into account Fermi
motion and Pauli blocking, the cross section already goes down by more
than 10\%. With the full model that also includes 
the $\Delta$ resonance in-medium modification, the reduction is  of the
order of 30\%. Furthermore, we corroborate the findings on nucleon targets 
(Fig.~\ref{fig:cs_nu}) about the  $N^*$ contributions 
[mostly the $N(1520)$] being sizable above 
$\sim 1.5$~GeV, specially for antineutrino cross sections. 

In the right-hand plots of Fig.~\ref{fig:incoh1}, we compare our results
with the predictions of Ref.~\cite{Zhang:2012xn}. As in the nucleon 
case (Fig.~\ref{fig:delta_term}), we
focus on the dominant $ \Delta P$ contribution and use the same $C_5^A(0)=1.14$ 
value and photon energy cut (200 MeV) as in Ref.~\cite{Zhang:2012xn}. 
When all the nuclear corrections are neglected,
we certainly obtain the same curves as in Fig.~\ref{fig:delta_term},
but multiplied by the number of nucleons (12). As can be observed in the figure, 
we find an excellent agreement both for neutrino and antineutrino cross sections. 
However, nuclear medium effects turn out to be much more important, leading to a much
larger suppression ($\sim 50\%$), in the calculation of
Ref.~\cite{Zhang:2012xn} for neutrinos. 
This seems surprising, first, because at this
moderately high neutrino energies, similar nuclear corrections should be obtained with 
both models. In particular, one would not expect significant differences in 
the $\Delta$ resonance broadening in the medium when calculated with Eq.~(\ref{eq:delta-self}) 
or with the spreading potential of Ref.~\cite{Zhang:2012xi}.
\footnote{We should mention that we agree better with the  $\Delta P$ cross section of 
Ref.~\cite{Zhang:2012xn} for neutrinos if we take an imaginary part of the
  $\Delta$ selfenergy twice bigger than the one in Eq.~(\ref{eq:delta-self}).}
Because of the larger nuclear suppression, the $\Delta P$ cross section 
found in Ref.~\cite{Zhang:2012xn} is smaller than the one obtained here 
in spite of the 14\% larger $C_5^A(0)$. In the antineutrino cross sections, the
difference is not so large, and  the medium effects shown in 
Ref.~\cite{Zhang:2012xn} are only slightly greater than those found
in the present work. As a consequence of the large reduction of the $\Delta P$ 
contribution on $^{12}$C, the contact terms become relatively important 
from $E_\nu =1$~GeV on, rapidly increasing and turning dominant 
above  1.5 GeV (see Fig. 3 of Ref.~\cite{Zhang:2012xn}). 
Indeed, contact terms compensate the suppression of the $ \Delta P$ mechanism, 
so that the  incoherent cross sections predicted in Ref.~\cite{Zhang:2012xn} 
are comparable to ours in the 1~GeV region, but become about 40\% (70\%) 
larger than our results for 2~GeV neutrinos (antineutrinos) even though 
the contributions from resonances heavier that the $\Delta$ were not 
 taken into account.

\begin{figure}[htb]
\begin{center}
\makebox[0pt] {\includegraphics[width=0.33\textwidth]{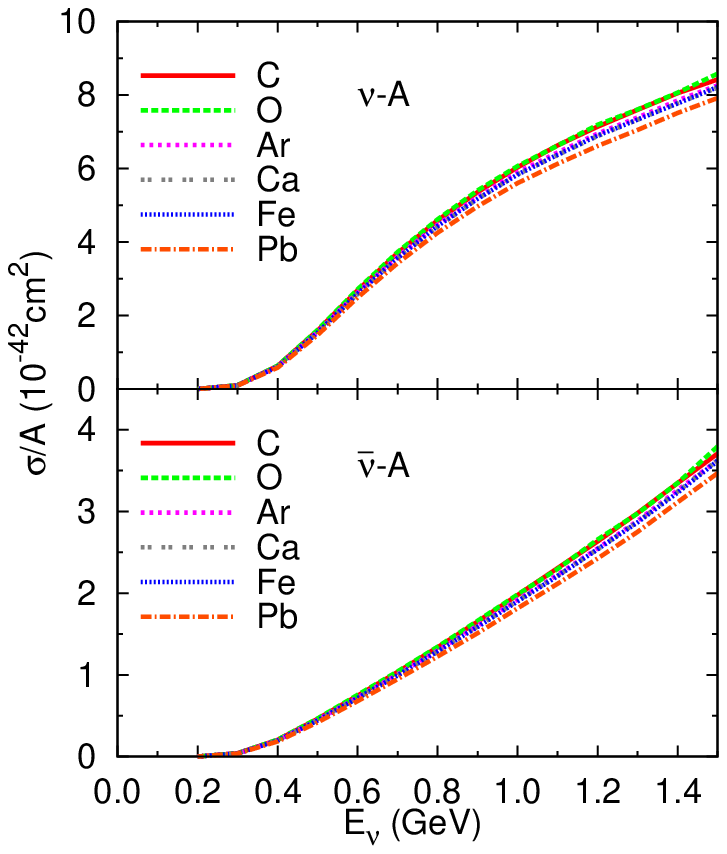}
\includegraphics[width=0.33\textwidth]{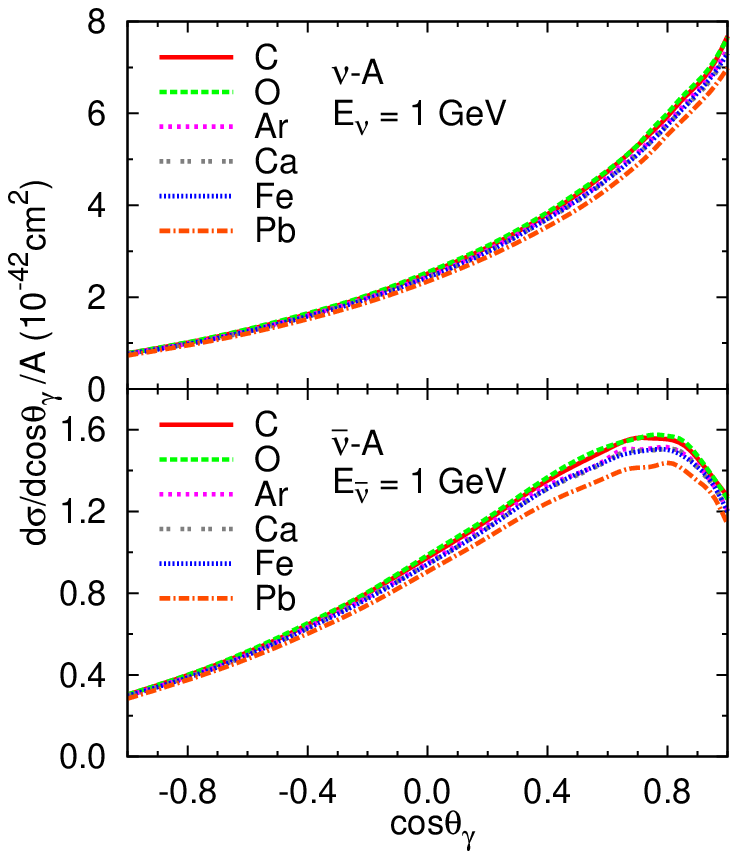}\includegraphics[width=0.33\textwidth]{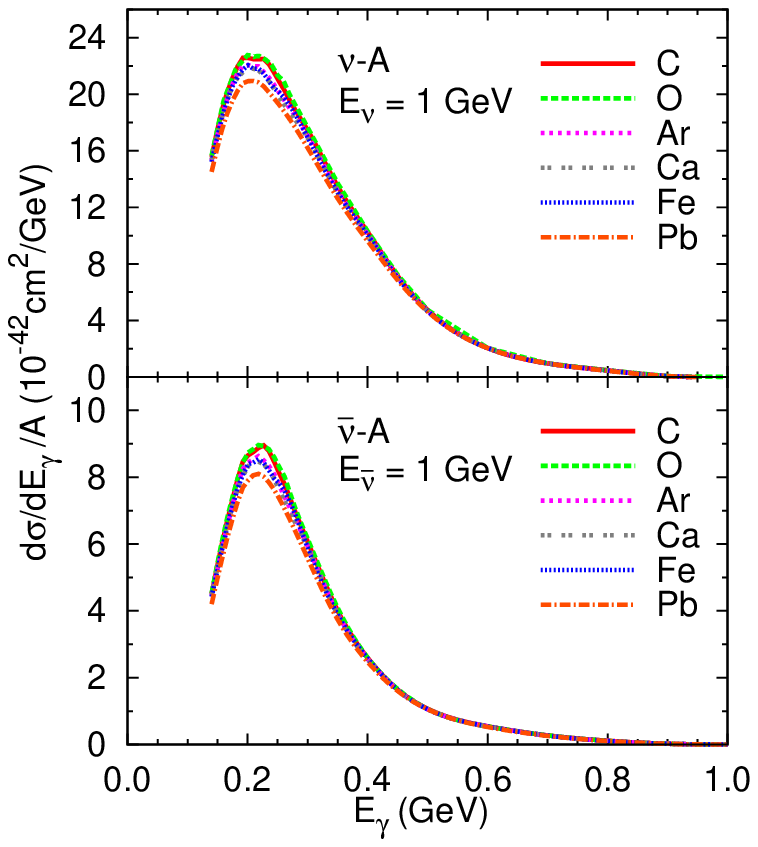}}
\caption{(color online) Neutrino (top) and antineutrino (bottom)
  incoherent NC$\gamma$ total cross sections as a function of the
  (anti)neutrino energy (left panels), photon angular (middle
  panels) and photon energy (right panels) 
  differential distributions at $E_{\nu,\bar\nu}=$1 GeV. The angle
  $\theta_\gamma$ is referred to the direction of the incoming
  (anti)neutrino beam. Results for
  different nuclei ($^{12}$C,$^{16}$O,$^{40}$Ar, $^{40}$Ca,$^{56}$Fe
  and $^{208}$Pb) divided by the number of nucleons are shown.  
All results are obtained with the full model, including nuclear effects and implementing an  $E_\gamma \geq 140$~MeV cut.}
\label{fig:incoh2}
\end{center}
\end{figure}
In Fig.~\ref{fig:incoh2}, we show total NC$\gamma$ incoherent cross
sections for different nuclei (carbon, oxygen, argon, calcium, iron
and lead) as a function of the (anti)neutrino energy. We also display
photon angular and energy distributions for an incoming (anti)neutrino
energy of 1 GeV. We notice the approximated $A-$scaling present in the 
results, which implies a mild $A$ dependence of nuclear effects. 
Nevertheless, the cross section is smaller for heavier nuclei, 
particularly $^{208}$Pb. We should stress that the observed deviation from scaling 
cannot be explained only by neutron cross sections being smaller than proton ones 
(around 15-20\% at $E_{\nu}\sim$ 1.5 GeV)\footnote{Note that the $\Delta$P contribution is the same on 
protons and neutrons. Thus, this dominant mechanism does not contribute to such differences.}.

Concerning the kinematics of the emitted photons, the main
features are similar to those in Figs.~\ref{fig:1gev} and
\ref{fig:2gev} for scattering on single nucleons. As in that case, 
the reaction is more forward for neutrinos than for
antineutrinos at $E_\nu =1$~GeV. 
In the outgoing photon energy distributions (right panels), 
the peak just above $E_\gamma=0.2$ GeV observed for nucleons  is 
reproduced here without any shift in the peak position but 
with slightly larger width as the target mass increases.

\subsubsection{Coherent reaction}
\label{sec:res2b}
 
\begin{figure}[h!]
\begin{center}
\includegraphics[width=0.5\textwidth]{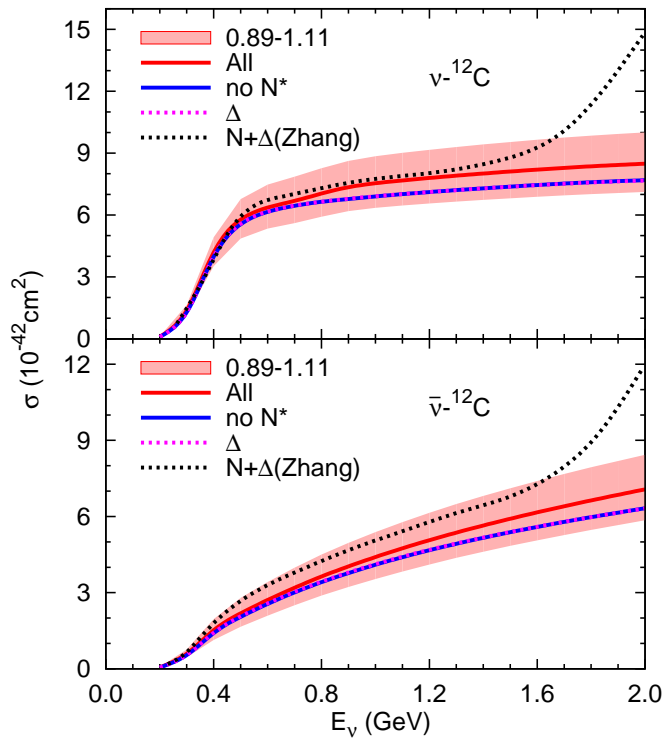}
\caption{(color online) Neutrino (top) and antineutrino (bottom) total
  NC$\gamma$ coherent cross sections on $^{12}$C, as a function of the
  (anti)neutrino energy. A photon energy cut of  $E_\gamma \geq 140$~MeV has 
been implemented. Red solid lines stand for
  results from the complete model derived in
  this work, including $\Delta$ resonance broadening, with error  bands 
determined by the uncertainty of $\pm 0.11$ in $C^A_5(0)$~\cite{Hernandez:2010bx}. 
The solid blue lines below, labeled as ``no $N^*$'', display the 
predicted cross sections without the $N^*$ amplitudes, while the magenta dotted ones
are the contributions from the $(\Delta P+ C\Delta P)$ mechanisms. 
We also show the predictions of Ref.~\cite{Zhang:2012xn} for nucleon and $\Delta$ mechanisms 
(red solid lines in Fig.4 of this reference).}
\label{fig:coh1}
\end{center}
\end{figure}

Total NC$\gamma$ coherent cross sections on 
carbon as a function of the (anti)neutrino energy are presented in
Fig.~\ref{fig:coh1}.
We display our results from the full calculation,
from $(\Delta P+ C\Delta P)$ alone, and without the mechanisms
from second $N^*$ resonance region.  
The $N^*$ contributions 
are quite small in the coherent channel, while the
$\Delta$ is absolutely dominant in both the neutrino and
the antineutrino modes. Nucleon-pole contributions are negligible 
because the coherent kinematics favors a strong cancellation between 
the direct and crossed terms of the amplitude. A similar effect has 
been observed in weak coherent pion production~\cite{AlvarezRuso:2007it}. 

For comparison, the predictions from the $(\Delta P+
C\Delta P+NP+CNP)$ part of the model of Ref.~\cite{Zhang:2012xn} 
are also plotted. They are slightly above our corresponding results 
(without $N^*$), and within the uncertainty band of our full-model curve, 
up to (anti)neutrino energies of 1.4--1.5 GeV. Above these energies, 
there is a a change of slope and a pronounced enhancement~\cite{Zhang:2012xn}. 
Moreover, in the model of this reference,
 the cross section above $E_{\nu,\bar\nu}=0.65$ GeV is not dominated by
the $(N+\Delta)$ mechanism, but by contact terms from higher order effective 
Lagrangians whose extrapolation to higher energies is uncertain. 
Indeed, for some choices of  parameters, coherent cross sections as large
as $25\times 10^{-42}$ cm$^2$ were obtained
for $E_{\nu,\bar\nu}=1.5$ GeV~\cite{Zhang:2012xn}. 
This amounts to a factor 3-4 larger
than our predictions. We should remind here that below 500 MeV, the
contact terms in the nucleon amplitudes are very small as
expected based on the power counting established in
Ref.~\cite{Serot:2012rd}. Because of the substantial reduction of the
$\Delta$ mechanisms, the contact terms in Ref.~\cite{Zhang:2012xn}
acquire further relevance when the processes take place in nuclei,
specially for the coherent reaction.

\begin{figure}[h!]
\begin{center}
\makebox[0pt] {\includegraphics[width=0.34\textwidth]{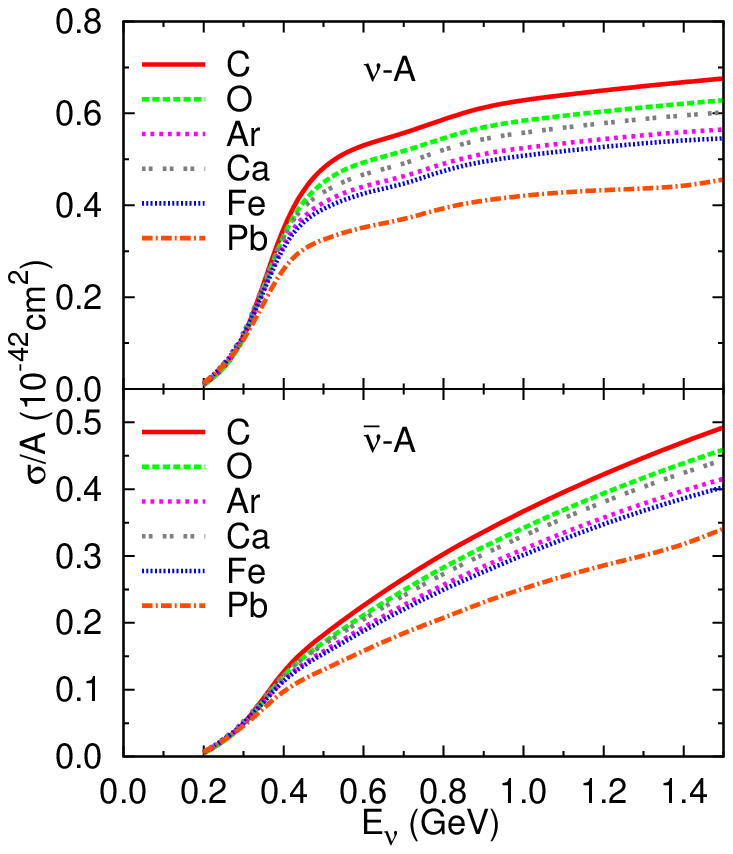}
\includegraphics[width=0.34\textwidth]{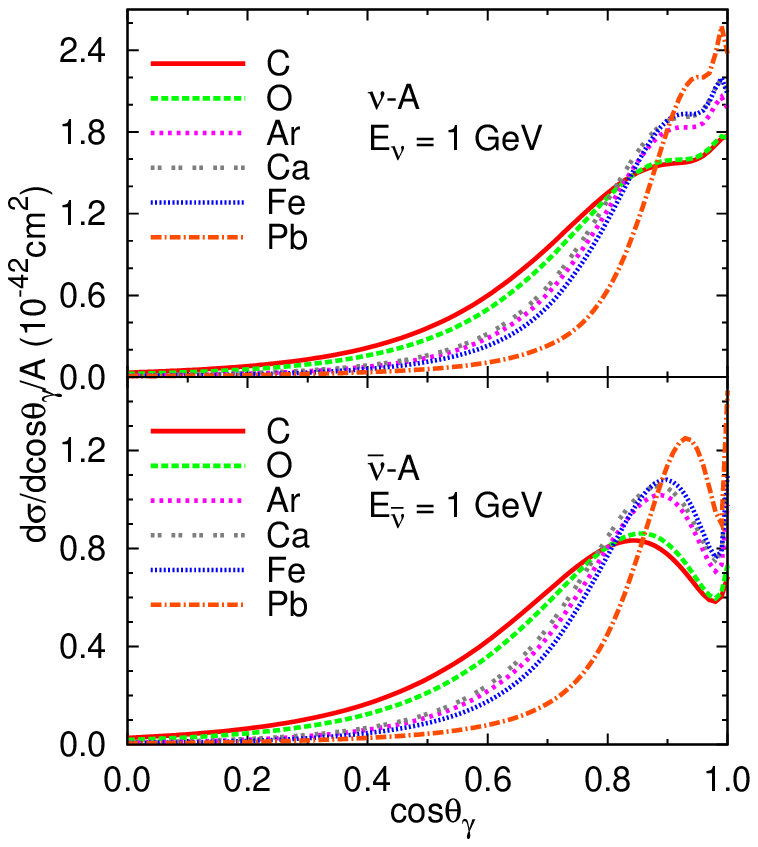}\includegraphics[width=0.34\textwidth]{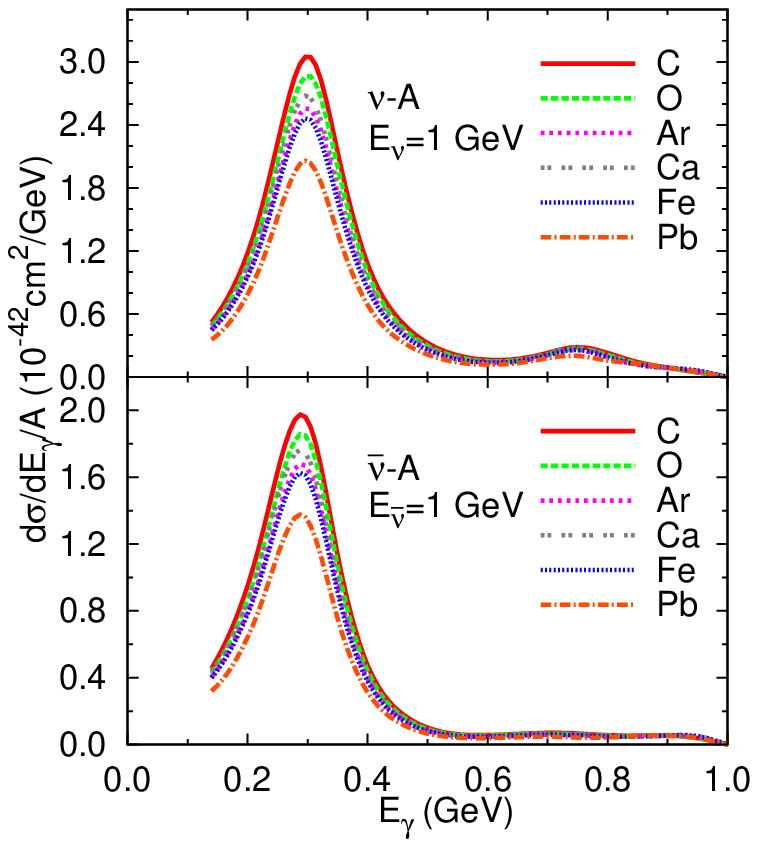}}
\caption{(color online) Neutrino (top) and antineutrino (bottom) total
cross sections (left panels)  photon angular (middle
  panels) and photon energy (right panels) differential distributions 
for the coherent NC$\gamma$ reaction, obtained with our full model. 
 The angle $\theta_\gamma$ is referred to the direction of the incoming
  (anti)neutrino beam. The kinematic region of $E_\gamma < 140$~MeV has been cut out.
Results for  different nuclei ($^{12}$C,$^{16}$O,$^{40}$Ar, $^{40}$Ca,$^{56}$Fe
  and $^{208}$Pb) divided by the number of nucleons are shown.}
\label{fig:coh2}
\end{center}
\end{figure}

Our results for coherent NC$\gamma$ total and differential cross sections on 
different nuclei are shown in Fig.~\ref{fig:coh2}. 
Neutrino (antineutrino) coherent cross sections are about a factor 15
 (10) smaller than the incoherent ones given in Fig.~\ref{fig:incoh2}. 
Thus, the relative relevance
 of the coherent channel with respect to the incoherent channel is
 comparable, if not greater than in the pion production reactions
 induced by neutrinos and antineutrinos, where it is of the order of
 few per cent~\cite{Hernandez:2010jf,Hernandez:2013jka}.  Notice that in 
these latter reactions the coherent cross section is 
further reduced (by around a factor
of two) because of the strong distortion of the outgoing pion, which is 
not present in photon production. It is also true that the incoherent cross section 
is reduced ($\sim 20-30$\%) by final state interactions, again absent for photons.

The coherent cross sections neither scale with $A$, like the incoherent one approximately does, nor with
$A^2$ as one would expect from the coherence of the dominant isoscalar
$\Delta P$ mechanism (sum of neutron and proton amplitudes). This is due to the presence
of the nuclear form factor (Fourier transform of the nuclear density
for momentum $\vec{q}-\vec{k}_\gamma$), see the first paragraph of Sec.~\ref{sec:coherent} 
and Eq.~(\ref{eq:Jmunu2}).  The nuclear form
factor gets its maximum values when $\vec{q} = \vec{k}_\gamma$, which corresponds to $q^2 = 0$. 
In this forward kinematics, the lepton tensor $L_{\mu\sigma}^{(\nu,\bar\nu)} \sim q_\mu q_\sigma$, and the 
vector part of the amplitude squared is zero due to CVC. Furthermore, the axial contribution, which is purely
transverse $\sim (\vec{k}_\gamma \times \vec{q})$ also vanishes. Therefore, the largest differential 
cross sections arise in kinematics that optimize the product of the amplitude squared of the elementary process
times the nuclear form factor. Such a balance also appears in 
the $(^3{\rm He},^3{\rm H}\,\pi^+)$ reaction on nuclear targets~\cite{FernandezdeCordoba:1992ky}  or 
in electron and photon induced reactions, making the electromagnetic coherent pion
production cross section a rather small fraction of the total inclusive nuclear absorption
one~\cite{Carrasco:1991we,Hirenzaki:1993jc}.  

The described pattern  strongly influences the photon angular
dependence of this reaction shown in the middle panels of
Fig.~\ref{fig:coh2} although in a non-trivial way because the 
$\theta_\gamma$ angle is given with respect to the direction of 
the incoming (anti)neutrino beam;  it is not the angle
formed by $\vec{q}$ and $\vec{k}_\gamma$, which is not observable. 
Actually, for each value of $\theta_\gamma$,  and integration over all possible $\vec{q}$ is
carried out. The details of the angular distributions are determined by interferences 
between the dominant $\Delta P$ mechanism and the $C\Delta P$ and $N(1520)$ ones, enhanced by the 
kinematic constrains imposed by the nuclear form factor. The impact of the latter is apparent 
in the width of the angular distributions which are narrower for heavier nuclei. 

Finally, in Fig.~\ref{fig:coh2} we display the outgoing photon energy
distributions (right panels).  In the coherent NC$\gamma$ reaction,
there are two massless particles in the final state, and a third one (the nucleus) 
which is very massive and has a small (negligible) kinetic energy
but can carry large momenta. The prominent peak observed for all nuclei is 
due to the dominant $\Delta$ 
resonance~\footnote{The energy of the resonant photons in LAB can be estimated from 
$M_R^2 \approx (k_\gamma + p^\prime)^2$. Taking $p^\prime$ from Eq.~(\ref{eq:pmu}) and for 
the situation $\vec{k}_\gamma \approx \vec{q}$ favored by the nuclear form factor, one 
finds that $k_{\gamma(R)}^0 \approx (M_R^2 - M^2)/(2 M)$. This gives 340~MeV for the $\Delta(1232)$ 
and 760~MeV for the $N(1520)$.} 
shifted to slightly lower 
invariant masses mostly by the energy dependence of the $\Delta$ width and the interference 
with the $C\Delta P$ mechanism. The peak position does not change appreciably from nucleus to nucleus, 
but it gets broader as $A$ increases. The second, smaller and broader peak that can be 
discerned for neutrinos but not for antineutrinos corresponds to the excitation of the $D_{13}(1520)$ 
resonance.

\section{Conclusions}
\label{sec:concl}

Neutral current photon emission on nucleons and nuclei
at intermediate energies has been theoretically investigated. 
We have developed a microscopic model for these reactions, 
in line with previous work on weak 
pion production~\cite{Hernandez:2007qq,AlvarezRuso:2007it, Amaro:2008hd,Hernandez:2013jka}.
We have critically reviewed previous models for the
NC$\gamma$ reaction on single nucleons~\cite{Zhang:2012xn,Hill:2009ek,Serot:2012rd} and 
nuclei~\cite{Zhang:2012xn,Zhang:2012aka,Zhang:2012xi} and compared our
results with those found in these references. From such a comparison, we have identified
some aspects of the above studies that either needed to be improved or
that were sources of uncontrolled systematic corrections.

NC$\gamma$ processes are important backgrounds for $\nu_\mu \to \nu_e$
and $\bar\nu_\mu \to \bar \nu_e$ appearance oscillation experiments
when photons are misidentified as $e^\pm$ from CCQE 
scattering of $\nu_e (\bar{\nu}_e)$.  At the relevant energies for
MiniBooNE and T2K experiments, the reaction is dominated by the 
weak excitation of the $\Delta(1232)$ resonance and its subsequent 
decay into $N\gamma$. Besides, we have also considered
non-resonant amplitudes that, close to threshold, 
are fully determined by chiral symmetry, and those driven by 
nucleon excited states from the second resonance region. Among 
the latter ones, we have found a sizable contribution of the $D_{13}(1520)$ state
for (anti)neutrino energies above 1.5 GeV. 

The model on the nucleon is extended to nuclear targets taking into account Fermi motion,
 Pauli blocking and the in-medium modifications of the $\Delta$ properties in a local 
Fermi gas, with Fermi momenta determined from proton and neutron density distributions. 
We have predicted different observables for several nuclei, including some of the 
common ones in current and future neutrino detectors (carbon, oxygen, argon, iron). 
The importance of nuclear corrections  in both the coherent and
incoherent channels has been stressed. The $A$ dependence of the cross section, which is 
different for the coherent and incoherent reactions, has also been discussed.  

In the light of our results, a new analysis of the NC induced photon production at MiniBooNE with the 
present model, aiming at the clarification of the role played by NC$\gamma$ events in the low-energy 
excess observed in this experiment, looks timely and important. It will be the subject of future research.

\appendix
\section{Relations between electromagnetic form factors and helicity amplitudes}
\label{sec:heliamp}

The $\gamma N \rightarrow R$ helicity amplitudes describe the
nucleon-resonance transition depending on the polarization of the
incoming virtual photon and the baryon-spin projections onto the direction of the 
photon momentum. We follow the definitions adopted in the MAID analysis~\cite{Drechsel:2007if,MAID}, 
from which the empirical parametrizations of the helicity amplitudes are taken. Namely\footnote{It should be pointed out that 
the $1/(\sqrt{2 M}\sqrt{2 M_R})$ factor in the definition of the helicity amplitudes comes from the normalization of Dirac spinors 
($\bar{u} u = 2 M$, $\bar{u}_R u_R = 2 M_R$) adopted in the present work.}
\begin{eqnarray}
&&{\cal A}_{1/2}=\sqrt{\frac{2\pi \alpha}{k_R}}
\left \langle S_z^*=\frac{1}{2}\, \left |\epsilon^{(+)}_{\mu}J^{\mu}_{EM}\right|\,S_z=-\frac{1}{2}\right\rangle \frac{1}{\sqrt{2 M}\sqrt{2 M_R}}  ,\\
&&{\cal A}_{3/2}=\sqrt{\frac{2\pi \alpha}{k_R}}
 \left \langle  S_z^*=\frac{3}{2}\, \left |\epsilon^{(+)}_{\mu}J^{\mu}_{EM}\right|\,S_z=\frac{1}{2}\right\rangle \frac{1}{\sqrt{2 M}\sqrt{2 M_R}},\\
&&{\cal S}_{1/2}=-\sqrt{\frac{2\pi \alpha}{k_R}}
\left \langle  S_z^*=\frac{1}{2}\, \left |\frac{|\vec{k}\,|}{\sqrt{Q^2}}
\epsilon^{(0)}_{\mu}J^{\mu}_{EM}\right|\,S_z=\frac{1}{2}\right\rangle \frac{1}{\sqrt{2 M}\sqrt{2 M_R}} ,
\end{eqnarray}
in the resonance rest frame (notice that $S_{1/2}$ is not Lorentz invariant) and with the $z$-axis parallel to the 
photon momentum. In other words,
\begin{equation}
k^\mu = (k^0,0,0,|\vec{k}|), \quad  p^\mu = (\sqrt{M^2+\vec{k}^2},0,0,-|\vec{k}|), \quad p^{*\mu} 
=  (p+k)^\mu =(M_R,0,0,0) 
\end{equation}
are the virtual photon, nucleon and resonance four-momenta. In addition, $Q^2 = - k^2$ and 
\begin{equation}
k_R = \frac{M^2_R-M^2}{2M_R} \,.
\end{equation}
The photon polarization vectors are given by
\begin{equation}
\epsilon_{(\pm)}^\mu = \mp \frac{1}{\sqrt{2}} (0,1,\pm i,0), \quad 
\epsilon_{(0)}^\mu = \frac{1}{\sqrt{Q^2}}(|\vec{k}\,|,0,0,k^0) \,.
\end{equation}
Finally, $S_z$ ($S_z^*$) denotes the nucleon (resonance) spin projection onto the $z$ axis.

With these definitions and the currents of Section~\ref{sec:gamma_amp}, it is straightforward 
to derive the following equations connecting helicity amplitudes and electromagnetic 
form factors~\cite{Leitner:2008ue}.

\paragraph{$N(1440)$}

\begin{eqnarray}
A^{p,n}_{1/2} &=& \sqrt{\frac{\pi \alpha [(M_R-M)^2+Q^2]}{2 M(M^2_R-M^2)}}
\left[\frac{Q^2}{2M^2}F^{p,n}_1 + \frac{M_R+M}{M}F^{p,n}_2 \right] \\[0.2cm]
S^{p,n}_{1/2} &=& -\sqrt{\frac{\pi \alpha [(M_R+M)^2+Q^2]}{M(M^2_R-M^2)}}
\frac{(M_R-M)^2+Q^2}{4M M_R} 
\left[\frac{M_R+M}{2M}F^{p,n}_1 - F^{p,n}_2 \right] 
\end{eqnarray}

\paragraph{$N(1535)$}

\begin{eqnarray}
A^{p,n}_{1/2} &=& \sqrt{\frac{\pi \alpha [(M_R+M)^2+Q^2]}{2 M(M^2_R-M^2)}}
\left[\frac{Q^2}{2M^2}F^{p,n}_1 + \frac{M_R-M}{M}F^{p,n}_2 \right] \\[0.2cm]
S^{p,n}_{1/2} &=& \sqrt{\frac{\pi \alpha [(M_R-M)^2+Q^2]}{M(M^2_R-M^2)}}
\frac{(M_R+M)^2+Q^2}{4M M_R} \left[\frac{M_R-M}{2M}F^{p,n}_1 - F^{p,n}_2 \right] 
\end{eqnarray}

\paragraph{$\Delta(1232)$}

\begin{eqnarray}
A^{p,n}_{1/2} &=& 
  \sqrt{\frac{\pi \alpha [(M_R-M)^2+Q^2]}{3M (M^2_R-M^2)}} \nonumber \\
&\times&  
\left[\frac{M^2+MM_R+Q^2}{M M_R}C^{V}_3 -\frac{M^2_R-M^2-Q^2}{2M^2} C^{V}_4 -\frac{M^2_R-M^2+Q^2}{2M^2}C^{V}_5 \right]
\\[0.2cm]
A^{p,n}_{3/2}  &=& 
 \sqrt{\frac{\pi \alpha [(M_R-M)^2+Q^2]} {M(M^2_R-M^2)}}  
  \left[\frac{M+M_R}{M}C^{V}_3 +\frac{M^2_R-M^2-Q^2}{2M^2} C^{V}_4+ \frac{M^2_R-M^2+Q^2}{2M^2} C^{V}_5 \right]
\\[0.2cm]
S^{p,n}_{1/2} &=& 
  \sqrt{\frac{\pi \alpha[(M_R+M)^2+Q^2]}{6M (M^2_R-M^2)}}  \frac{(M_R-M)^2+Q^2}{M^2_R} 
\left[\frac{M_R}{M}C^{V}_3 + \frac{M_R^2}{M^2} C^{V}_4
+ \frac{M^2_R+M^2+Q^2}{2M^2} C^{V}_5 \right]
\end{eqnarray}

\paragraph{$N(1520)$}

\begin{eqnarray}
A^{p,n}_{1/2} &=& \sqrt{\frac{\pi \alpha [(M_R+M)^2+Q^2]}{3M (M^2_R-M^2)} } \nonumber \\
&\times& \left[\frac{M^2-MM_R+Q^2}{M M_R}C^{p,n}_3 -\frac{M^2_R-M^2-Q^2}{2M^2} C^{p,n}_4 
-\frac{M^2_R-M^2+Q^2}{2M^2} C^{p,n}_5 \right] \label{eq:a121520}
\\ [0.2cm]
A^{p,n}_{3/2} &=& \sqrt{\frac{\pi \alpha [(M_R+M)^2+Q^2]}{M(M^2_R-M^2)} } 
\left[\frac{M-M_R}{M}C^{p,n}_3 -\frac{M^2_R-M^2-Q^2}{2M^2} C^{p,n}_4
-\frac{M^2_R-M^2+Q^2}{2M^2} C^{p,n}_5 \right] \label{eq:a321520}
\\ [0.2cm]
S^{p,n}_{1/2} &=& -\sqrt{\frac{\pi \alpha [(M_R-M)^2+Q^2]}{6M(M^2_R-M^2)} } 
\frac{(M_R+M)^2+Q^2}{M^2_R} 
\left[\frac{M_R}{M}C^{p,n}_3 +\frac{M_R^2}{M^2} C^{p,n}_4
+\frac{M^2_R+M^2+Q^2}{2M^2} C^{p,n}_5 \right] \label{eq:s121520}
\end{eqnarray}

\section{Off diagonal Goldberger-Treiman relations }
\label{sec:axialcoupling}

We consider an effective Lagrangian for the $R N\pi$ vertex, 
which is then used to calculate the  $\pi N$ decay
width of the resonance. Using the Particle Data Group (PDG)~\cite{Beringer:1900zz} 
values for the decay width and $\pi N$ branching ratio, one can fix the  $RN\pi$ coupling.
Thanks to PCAC
\begin{equation}
\partial_\mu A^\mu_{NCI} (x) = 2 f_\pi m^2_\pi \pi^0 \,,
\end{equation}
the latter coupling can be related to the dominant axial coupling in 
$A^\mu_{NCI}$, which is the isovector part of the neutral current.
This is  the so called off diagonal GT relation.  
It establishes that in the soft pion limit
\begin{equation}
p_{\pi^0}^\mu \left \langle R | A^{NCI}_{\mu} (0) | N \right \rangle  =
- 2if_\pi  \left \langle R | {\cal L}_{RN\pi} | N \pi^0 \right \rangle  
\end{equation}

As in Refs.~\cite{Leitner:2008ue,Leitner:2009zz}, we distinguish between different cases, depending on the
spin, parity and isospin of the resonance. Let us start with spin 1/2 states with isospin 1/2, like the
$P_{11}(1440)$ and $S_{11}(1535)$. In this case
\begin{equation}
 {\cal L}_{R_{1/2}N \pi} = \frac{f}{m_{\pi}}  \bar \Psi
\left\lbrace \begin{array}{c}  \gamma^\mu \gamma_5  \\ \gamma^\mu \end{array} \right\rbrace
\left (\partial_\mu  {\vec\pi} \cdot \vec{\bf \tau} \right) \Psi_{R_{1/2}}  + h.c. \label{eq:N*Npi} \,,
\end{equation}
where $\Psi$, $\Psi_{R_{1/2}}$ and $\vec{\pi}$ are the nucleon, resonance and pion fields\footnote{Our convention is such that $(\pi^1 - i \pi^2)/\sqrt{2}$ creates a $\pi^-$ or annihilates a $\pi^+$ while a $\pi^3 = \pi^0$ field creates or annihilates a $\pi^0$.}; 
$\vec{\bf \tau}$ are the isospin Pauli matrices. The upper (lower) Lagrangian 
holds for positive (negative) parity resonances. The partial $R \to \pi N$ decay width is 
\begin{equation}
\Gamma_{R_{1/2} \to N \pi} =\frac{3}{4\pi M_R} \left(\frac{f}{m_\pi}\right)^2  
(M_R \pm M)^2 \left(E_N \mp M\right) |\vec{p}_N| \,,
\label{eq:width_J12}
\end{equation}
where 
\begin{equation}
E_N =
\sqrt{M^2+\vec{p}_N^{\,2}}= 
\frac{M^2_R + M^2 - m^2_\pi}{2M_R}\,.
\end{equation}
The upper (lower) signs in Eq.~(\ref{eq:width_J12}) stand for positive (negative) parity resonances. The off diagonal GT relation amounts to
\begin{eqnarray}
F_{A(R)}(0) = -2  \frac{f}{m_\pi} f_\pi
\end{eqnarray}
regardless of the parity. The coupling $F_{A(R)}(0)$ defined  in Eq.~(\ref{eq:faj12t12}) is now expressed in terms of 
 $f/m_\pi$ extracted from the $R \to \pi N$ decay width given above.

For $J=3/2$ resonances, $\Delta(1232)$ and $D_{13}(1520)$ in our case, 
\begin{equation}
{\cal L}_{R_{3/2}N \pi} = \frac{f^*}{m_{\pi}} \bar \Psi 
\left\lbrace \begin{array}{c} 1  \\ \gamma_5 \end{array} \right\rbrace
\left (\partial_\mu  {\vec\phi} \cdot \vec{\bf t} \right) \Psi^\mu_{R_{3/2}}  + h.c. \label{eq:N*s32Npi}
\end{equation}
where $\Psi^\mu_{R_{3/2}}$ is the resonance spin 3/2 field in the Rarita-Schwinger representation; 
$\vec{\bf t} = \vec{\bf \tau}$ stands for isospin 1/2 resonances and $\vec{\bf t} = \vec{\bf T}$ 
($3/2$ to $1/2$ isospin transition operator)\footnote{Normalized in such a way that the isospin matrix element $ \left \langle \frac{3}{2} \frac{3}{2} \right | T_1^\dagger + i T_2^\dagger \left|\frac{1}{2}\frac{1}{2} \right \rangle = - \sqrt{2}$.}  for isospin 3/2 ones. The upper (lower) Lagrangian 
applies for positive (negative) parity states. The partial $R \to \pi N$ decay width is then given by
\begin{equation}
\Gamma_{R_{3/2} \to N \pi} =\frac{c_I}{6\pi} \left(\frac{f^*}{m_\pi}\right)^2  
\frac{E_N \pm M}{2M_R} |\vec{p}_N|^3,
\label{eq:width_J32}
\end{equation}
where the upper (lower) sign stands for positive (negative) parity resonances while $c_I= 1 (3)$ for isospin 1/2 (3/2). Then we deduce 
\begin{eqnarray}
C^A_{5(R)}(0) = d_I  \frac{f^*}{m_\pi} f_\pi \,,
\label{eq:GTJ3/2}
\end{eqnarray}
where the numerical value of $f^*/m_\pi$ is obtained from Eq.~(\ref{eq:width_J32}). The coefficient 
$d_I = -2$ is for isospin 1/2 states like the $D_{13}(1520)$ and $d_I = \sqrt{2/3}$ for isospin 3/2 ones, like 
the $\Delta(1232)$. The corresponding $C^A_{5(R)}(0)$ couplings determined by this GT relation were defined in 
Eqs.~(\ref{eq:c5q2N1520}) and (\ref{eq:c5}). It should be reminded that for the $N-\Delta(1232)$ transition, rather than 
the $C^A_{5}(0)$ value from Eq.~(\ref{eq:GTJ3/2}), we use the one fitted  in Ref.~\cite{Hernandez:2010bx}
to the $\nu_\mu p \to \mu^- p \pi^+$ ANL and BNL bubble chamber data.

\section{Decay modes of the second region resonances }
\label{sec:width}
In Table~\ref{tab:enes*}, we compile the most relevant
$P_{11}(1440)$, $D_{13}(1520)$ and $S_{11}(1535)$ decay modes and their branching 
ratios, taking values within the ranges of the PDG estimates ~\cite{Beringer:1900zz}. 
\begin{table}[htp]
\caption{Main decay modes, branching fractions
  ($\Gamma_i/\Gamma$) and relative angular momenta $L$ of the decay particles,  
for the $N^*$ resonances considered in this
  work.} \label{tab:enes*} \vspace{0.1cm}
\begin{center}
\begin{tabular}{cccc|cccc|ccc}
\hline\hline
\multicolumn{3}{c}{$N(1440)$} & & \multicolumn{3}{c}{$N(1520)$}& &
\multicolumn{3}{c}{$N(1535)$}  \tstrut\\ 
 Mode          & Fraction(\%) & $L$  &  &  Mode          &
 Fraction(\%) & $L$&  &  Mode          & Fraction(\%) & $L$  \\\hline
$N\pi$         & 65            & 1   &  & $N\pi$         & 60            & 2 &  &$N\pi$         & 45            & 0   \\
$\Delta\pi$    & 20            & 1   &  & $\Delta\pi$    & 15            & 0&  &$N\eta$        & 42            & 0   \\
$N\sigma$      & 15            & 0   &  & $\Delta\pi$    & 12.5          & 2&  &$\Delta\pi$    & 1             & 2   \\
               &               &     &  & $N\rho$        & 9           & 0&  &$N\rho$        & 2             & 0   \\
               &               &     &  & $N\rho$        & 3.5
 & 2&  &$N\sigma$      & 2             & 1   \\
               &               &     &  &         & 
 & &  &$N(1440)\pi$   & 8             & 0   \\
\hline\hline
\end{tabular}\end{center}
\end{table}

To obtain the partial width of a decay mode into
unstable particles we use~\cite{Buss:2008zz}  
\begin{equation}
\Gamma_{R \to ab}(W) = \Gamma_{R \to ab}(W=M_R) 
\frac{\rho_{ab}(W)}{\rho_{ab}(M_R)}
\end{equation}
where $W$ denotes the resonance invariant mass. 
The function $\rho_{ab}$ is given by
\begin{equation}
\rho_{ab}(W)= \int d(p^2_a) d(p^2_b) {\cal A}(p^2_a){\cal A}(p^2_b)
\frac{p^{2L+1}_{ab}(W^2,p_a^2,p_b^2)}{W}   \,\Theta(W-\sqrt{p_a^2}-\sqrt{p_b^2}) , \qquad p_{ab}^2 =
\frac{\lambda(W^2,p^2_a, p^2_b)}{4W^2} \,,
\end{equation} 
where $p_{ab}$ denotes the center-of-mass momentum of the final state products,
and $L$ the relative angular momentum (Table~\ref{tab:enes*}). 
The vacuum spectral function ${\cal A}_{a}$ reads 
\begin{equation}
{\cal A}(p^2_{a}) = -\frac{1}{\pi} \mathrm{Im} \left(\frac{1}{p^2_{a}-M^2_{a}+iM_{a}\Gamma_{a}(p^2_{a})} \right) \,.
\end{equation}
If one of the decay products ($a$) is a stable particle, 
then $\Gamma_a=0$ and the vacuum spectral function can be written as
\begin{equation}
{\cal A}(p^2_a)= \delta (p^2_a-M^2_a) \,
\end{equation}
so that $\rho_{ab}$ becomes,
\begin{equation}
\rho_{ab}(W) = \frac{M_b}{\pi W}\int d(p^2_b) \frac{\Gamma_b(p^2_b)}{(p^2_b-M^2_b)^2+M^2_b \Gamma^2_b(p^2_b)}\, p^{2L+1}_{ab}(W^2,M_a^2,p_b^2) \,\Theta(W- M_a -\sqrt{p_b^2}) \,.
\end{equation}
If both final particles are stable, then 
\begin{equation}
\rho_{ab}(W) = \frac{p^{2L+1}_{ab}(W^2,M_a^2,M_b^2)}{W} \,\Theta(W-M_a-M_b) \,.
\end{equation}

\begin{acknowledgments}

 We thank M.J. Vicente Vacas for helpful discussions and M. Valverde 
for collaboration in an early stage of this project. This research
 has been supported by the Spanish Ministerio de Econom\'\i a y
 Competitividad and European FEDER funds under the contracts
 FIS2011-28853-C02-01 and FIS2011-28853-C02-02 and the Spanish
 Consolider-Ingenio 2010 Program CPAN (CSD2007-00042), by
 Generalitat Valenciana under contract PROMETEO/2009/0090 and by the
 EU HadronPhysics3 project, grant agreement no. 283286.
\end{acknowledgments}

\bibliography{neutrinos}

\end{document}